\documentclass[journal]{IEEEtran}
\usepackage{amsmath}
\usepackage{float}
\usepackage{graphicx}
\usepackage{multirow}
\usepackage{amsfonts,amssymb}
\usepackage{cite}
\usepackage{color}
\usepackage{algorithm}
\usepackage{algorithmic}

\ifCLASSINFOpdf
\else
\fi
\begin{document}
%
\title{A Low Complexity Learning-based Channel Estimation for OFDM Systems with Online Training}

\author{\IEEEauthorblockN{Kai Mei,
Jun Liu,
Xiaoying Zhang,
Kuo Cao,
Nandana Rajatheva,
and Jibo Wei}

\thanks{(c) 2021 IEEE. Personal use is permitted, but republication/redistribution requires IEEE permission. See https://www.ieee.org/publications/rights/index.html for more information. Citation information for this article : DOI 10.1109/TCOMM.2021.3095198.

Kai Mei, Jun Liu, Xiaoying Zhang, Kuo Cao, and Jibo Wei are with the
College of Electronic Science and Technology, National University of Defense
Technology, Changsha 410073, China (E-mail: {meikai11, liujun15, zhangxiaoying, caokuo18, wjbhw}@nudt.edu.cn).

Nandana Rajatheva is with Center for Wireless Communications, University of Oulu, Oulu 90570, Finland (E-mail: nandana.rajatheva@oulu.fi).}}


%



\IEEEtitleabstractindextext{%
\begin{abstract}
In this paper, we devise a highly efficient machine learning-based channel estimation for orthogonal frequency division multiplexing (OFDM) systems, in which the training of the estimator is performed online. A simple learning module is employed for the proposed learning-based estimator. The training process is thus much faster and the required training data is reduced significantly. Besides, a training data construction approach utilizing least square (LS) estimation results is proposed so that the training data can be collected during the data transmission. The feasibility of this novel construction approach is verified by theoretical analysis and simulations. Based on this construction approach, two alternative training data generation schemes are proposed. One scheme transmits additional block pilot symbols to create training data, while the other scheme adopts a decision-directed method and does not require extra pilot overhead. Simulation results show the robustness of the proposed channel estimation method. Furthermore, the proposed method shows better adaptation to practical imperfections compared with the conventional minimum mean-square error (MMSE) channel estimation. It outperforms the existing machine learning-based channel estimation techniques under varying channel conditions.

\end{abstract}

}

\maketitle

\IEEEdisplaynontitleabstractindextext

%
\IEEEpeerreviewmaketitle

\section{Introduction}

\IEEEPARstart{O}{rthogonal} frequency division multiplexing (OFDM) has been widely used in wireless broadband systems. Channel estimation is of vital importance for coherent OFDM systems and has been extensively studied. In the systems that employ pilot-aided channel estimation, the subcarriers that carry pilot signals are multiplexed with the data subcarriers \cite{4202196}. Initial channel estimates at pilot subcarriers are often obtained first with least-squares (LS) estimation, and then interpolation schemes are exploited to produce the channel responses at data subcarriers. Many techniques can be adopted for interpolating and a favored one is the minimum mean-square error (MMSE) interpolation due to its excellent performance \cite{4267831}. To perform MMSE interpolation, one needs to know the second-order channel statistics. Therefore, additional manipulation is required for the acquisition of channel statistics, and the channel statistics mismatch may cause severe performance degradation.

Recently, deep learning (DL)-based channel estimation has been proposed for wireless communications systems including OFDM systems, multiple input multiple output (MIMO), and massive MIMO systems. DL techniques, such as deep neural network (DNN) \cite{8272484,8672767,8693948,8798971,KangDeep,8663988,GaoComNet,8933050,8813060,9048929,8949757}, recurrent neural network (RNN) \cite{8847452}, and convolution neural network (CNN) \cite{8640815,8353153,8933411,9131701,8902116}, have been leveraged to perform channel estimation. These methods merely need a dataset to train the neural network, and channel statistics are thus not required in the channel estimation phase. Furthermore, these DL techniques bring tangible new benefits. Specifically, DNN has demonstrated a surprising ability to capture the complicated characteristics of data and thus the DNN-based approach has advantages when wireless channels suffer from serious distortion and interference. CNN shines in image recovering and image denoising. Therefore, the CNN-based approach has a high potential to improve the estimation performance. The RNN with long short time memory (LSTM) has the ability to capture long-time information in the data sequence. Due to this, the RNN-based approach can track a time-varying channel dynamically by exploiting the behavior of correlation in the time domain.

Most of those DL-based approaches train the estimator in an offline manner, where a large-scale training dataset is required. It is cumbersome to create high-quality datasets that agree well with real-world applications and there is a shortage of available datasets for DL-based communication systems compared with image processing and natural language processing (NLP) applications. In addition, these approaches are not amenable to wireless systems where channels change rapidly. Although an untrained DNN-based estimator proposed in \cite{8949757} does not require offline training, the architecture of DNN in \cite{8949757} is designed for massive MIMO channel estimation, which cannot be directly extended to the OFDM systems. However, employing an online training scheme can avoid the aforementioned problems faced by the offline training manner \cite{8715649}. This is because, in an online training scheme, training data is collected during the real-time transmission and thus agrees well with real-world applications. Besides, since the learning module of the learning-based estimator is trained in real-time, it can be adapted to the changing channel conditions. However, little attention has been paid to the machine learning (ML)-based channel estimation with an online training scheme. In designing ML-based channel estimation with online training, there are three main challenges. First, in the existing ML-based channel estimation methods, the true channel responses are used as labels in training data. However, the true channel responses are not known in practice. Second, the size of the provided training data is usually limited, which may not be enough to train the learning module. Third, the training procedure should be completed fast to meet the latency constraint.

In this paper, we propose a linear machine learning (LML)-based channel estimation with online training. The main contributions are as follows.

\begin{itemize}
\item We devise a low complexity machine learning-based channel estimation with an online training scheme. In the proposed scheme, training data is generated during transmission and the estimator is trained periodically online. Therefore, the estimator can be optimized for the real-time channel conditions and adapt to channel changes.

\item We propose a novel training data construction approach. In the proposed scheme, the true channel responses are replaced by their estimates to label training data so that the training data can be collected online. The proposed training data structure is validated by theoretical analysis and simulations.

\item Based on the proposed data structure, two alternative training data generation schemes are given. In one scheme, adding one block pilot symbol into the conventional OFDM frame, the required training data can be provided using the LS estimation results of the block pilot symbol. The other scheme creates training data in a decision-directed manner, which improves the spectrum efficiency.

\item A linear structure is employed as the learning module, where the outputs are the linear combination of the inputs. A lower training period and a smaller amount of training data can be achieved due to the simplicity of the learning module. In addition, since the training data provided by the proposed generation scheme is limited, the linear structure shows better performance compared to the other machine learning techniques, e.g., DNN, which is demonstrated by simulation experiments.

\end{itemize}

The remainder of this paper is organized as follows. The system model, conventional channel estimation methods, and machine learning-based channel estimation are introduced in Section \ref{Sec.CE}. The proposed linear machine learning-based channel estimator is illustrated in Section \ref{sec.LMLCE}. The online training data generation schemes are explained in Section \ref{sec.Generation}. The performance of the proposed estimator and the proposed training data generation schemes are validated by simulations in Section \ref{Sec.Sim}. Section \ref{sec.Conclusion} concludes this paper.

Notation: We use boldface small letters and capital letters to denote vectors and matrices, respectively. $\mathbb{E}\left[  \cdot  \right]$, ${\left\|  \cdot  \right\|_2}$, and $\otimes$ represent the expectation, the Euclidean norm, and circular convolution, respectively. The superscripts ${\left(  \cdot  \right)^ * }$, ${\left(  \cdot  \right)^{\rm{T}}  }$, ${\left(  \cdot  \right)^{\rm{H}}  }$, ${\left(  \cdot  \right)^{ - 1}}$, and ${\left(  \cdot  \right)^ {\dagger} }$ denote the conjugate of complex, the transpose, the Hermitian transpose, the inversion, and Moore-Penrose (MP) generalized matrix inverse, respectively. The superscripts $\left(  \cdot  \right)^{\rm{t}}$ and $\left(  \cdot  \right)^{\rm{f}}$ stand for time domain and frequency domain, respectively.

\section{Channel Estimation}
\label{Sec.CE}

\subsection{System Model}
\label{Sec.model}

We consider a single-input single-output (SISO) OFDM system under a slowly fading wide-sense stationary uncorrelated scattering (WSSUS) channel, in which the channel impulse response (CIR) is constant during one OFDM symbol. The received signal in the time domain can be expressed as
\begin{equation}
\label{equ.timedomain}
{\bf{y}}^{\rm{t}} = {{\bf{x}}^{\rm{t}}} \otimes {{\bf{h}}^{\rm{t}}} + {\bf{z}},
\end{equation}
where ${{\bf{x}}^{\rm{t}}}$ is a vector containing the transmitted signal in time domain and ${{\bf{h}}^{\rm{t}}}$ is the CIR vector. ${\bf{z}}$ represents the white Gaussian noise vector. We assume that the length of the cyclic prefix (CP) is larger than the channel length and the time and frequency synchronizations are accurate. After the CP removal and discrete Fourier transform (DFT) operation, the received signal can be written as
\begin{equation}
\label{equ:chap2.28}
{\bf{y}}^{\rm{f}} = {{\bf{X}}^{\rm{f}}} {{\bf{h}}^{\rm{f}}} +\tilde {\bf{z}},
\end{equation}
where ${\bf{X}}^{\rm{f}}$ is a diagonal matrix containing the transmitted signal. ${\bf{h}}^{\rm{f}}$ denotes the channel frequency response (CFR) vector. $\tilde {\bf{z}}$ is a white Gaussian noise vector.

\begin{figure}
\begin{centering}
\includegraphics[scale=0.8]{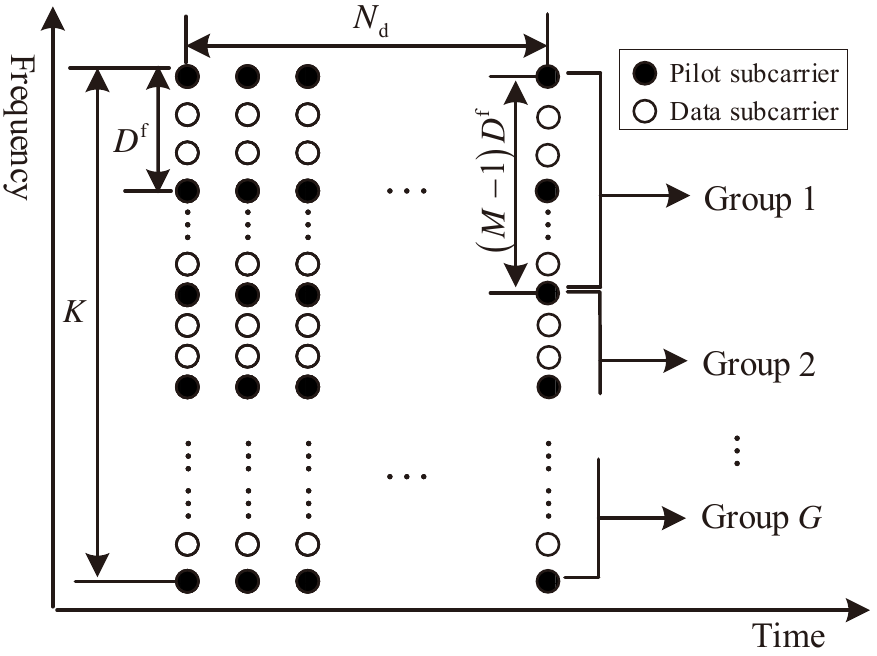}

\end{centering}
\caption{ Sketch diagram for the OFDM symbol partition.}
\label{Fig.DivSymbol}
\end{figure}

In this paper, we consider a low complexity channel estimation scheme, in which an OFDM symbol is divided into several groups and channel estimation is performed on each group individually. For simplicity, we assume that the same estimator is used to perform channel estimation for all the groups. To this end, we adopt the pilot arrangement shown in Fig. \ref{Fig.DivSymbol}, where the division of the OFDM symbol is also illustrated. $K$, $ D^{\rm f}$, $N_{\rm d}$, and $M$ represent the number of available subcarriers per one OFDM symbol, the spacing between pilot subcarriers (PS), the number of OFDM symbols or data symbols in a frame, and the number of pilot subcarriers in one group, respectively. We assume that $M \ll P$, where $P $ denotes the number of pilot subcarriers in an OFDM symbol. Since the computational complexity of interpolation is linear to $M$, a small $M$ is crucial for reducing the implementation complexity of the channel estimator. $G$ is the number of groups and $G=K/\left({(M-1)D^{\rm f}}\right)$.

In Fig. \ref{Fig.DivSymbol}, the pilots are regularly placed at the same locations for each OFDM symbol, which belong to the comb-type pilot arrangement \cite{4267831}. In addition, the subcarriers at both ends are pilot subcarriers so that the groups have the same data structure. Hence, the index set of the pilot subcarriers in an OFDM symbol is given by
\begin{equation}
\label{equ.PSindex}
\Omega \triangleq \left\{ {1,1+D^{\rm f},1+2 D^{\rm f},\cdots,K} \right\}.
\end{equation}

Since all the groups use the same estimator, we only consider the channel estimation problem for the $k{\rm th}$ group, specifically using ${{\bf{X}}_{{\rm p}\_k }^{\rm f}}$ to estimate $ {\bf{ h}}_{{\rm d}\_k}^{\rm{f}}$. ${{\bf{X}}_{{\rm p}\_k }^{\rm f}}$ is a $M \times M$ diagonal matrix containing the pilot signals in the $k{\rm th}$ group. $ {\bf{ h}}_{{\rm d}\_k}^{\rm{f}}$ is a $S \times 1$ vector containing the CFRs at data subcarriers in the $k{\rm th}$ group, where $S$ is the number of data subcarriers in a group.

Note the channel estimation method proposed in this paper is not only developed for the pilot arrangement depicted in (\ref{equ.PSindex}) but also applicable for the general case of the comb pilot arrangement. This is because we can simply repeat the procedures of the proposed method to generate several estimators when the subcarriers at both ends are not pilot subcarriers and more than one estimator may be required. This only leads to additional computation.

\subsection{Conventional Channel Estimation Methods}
\label{Sec.ConvMeth}

Conventionally, channel estimation for the considered pilot pattern is performed in two steps. The CFRs at the pilot subcarriers are first obtained and then, interpolation is applied to estimate the CFRs at the data positions.

The LS method is usually employed to estimate the CFRs at the pilot positions, which is performed as
\begin{equation}
\label{equ.LS}
{{{\bf{\hat h}}}_{{\rm p}\_k }^{\rm f}} = \left({{\bf{X}}_{{\rm p}\_k }^{\rm f}}\right)^{ - 1}{{\bf{y}}_{{\rm p}\_k }^{\rm f}},
\end{equation}
where ${{\bf{y}}_{{\rm p}\_k }^{\rm f}}$ is a vector containing the received signals corresponding to ${{\bf{X}}_{{\rm p}\_k }^{\rm f}}$ in frequency domain.

These LS estimates are then interpolated to estimate $ {\bf{ h}}_{{\rm d}\_k}^{\rm{f}}$. The process of the interpolation can be denoted as
\begin{equation}
\label{equ.Int}
{\hat {\bf{ h}}_{{\rm d}\_k}^{\rm{f}}} = {{\bf{W}}_{\rm{d}}} {{{\bf{\hat h}}}_{{\rm p}\_ k}^{\rm f}},
\end{equation}
where ${{\bf{W}}_{\rm{d}}}$ is the $S \times M$ interpolation matrix. Denote ${{{\bf{\hat h}}}_{{\rm{d}}}^{\rm{f}}} = [{{{{\bf{\hat h}}}_{{\rm{d\_1}}}^{\rm{f}}},\cdots,{{{\bf{\hat h}}}_{{\rm d}\_G}^{\rm{f}}} }]^{\rm T}$ as the vector containing the CFRs at data positions in an OFDM symbol.

The conventional interpolation techniques reviewed in \cite{4267831} can be regarded as different approaches to calculate ${{\bf{W}}_{\rm{d}}}$. Among those techniques, the optimum one is the MMSE interpolation \cite{4267831}, in which ${{\bf{W}}_{\rm{d}}}$ is given by
\begin{equation}
{{\bf{W}}_{\rm{d\_MMSE}}} = {\bf R}_{{{{\bf{ h}}_{{\rm d}\_k}^{\rm{f}}} {{{\bf{ h}}}_{{\rm p}\_ k}^{\rm f}}}}\left ({{\bf R}_{{{{\bf{ h}}}_{{\rm p}\_ k}^{\rm f}}{{{\bf{ h}}}_{{\rm p}\_ k}^{\rm f}}}+\sigma_n^2 {\bf I}}\right )^{-1},
\label{equ.MMSE}
\end{equation}
where ${{\bf{ h}}}_{{\rm p}\_ k}^{\rm f}$ contains the actual channel responses for ${\hat{\bf{ h}}}_{{\rm p}\_ k}^{\rm f}$. ${\bf R}_{{{\bf{ h}}_{{\rm d}\_k}^{\rm{f}}} {{{\bf{ h}}}_{{\rm p}\_ k}^{\rm f}}} = \mathbb{E}\left[ {{{\bf{ h}}_{{\rm d}\_k}^{\rm{f}}}{\left( {{{\bf{ h}}}_{{\rm p}\_ k}^{\rm f}} \right)}^{\text{H}}} \right]$ is the channel correlation matrix between the data subcarriers and pilot subcarriers. ${\mathbf{R}}_{{{{\bf{ h}}}_{{\rm p}\_ k}^{\rm f}}{{{\bf{ h}}}_{{\rm p}\_ k}^{\rm f}}} = \mathbb{E}\left[ {{{{\bf{ h}}}_{{\rm p}\_ k}^{\rm f}}{{\left( {{{\bf{ h}}}_{{\rm p}\_ k}^{\rm f}} \right)}^{\text{H}}}} \right]$ denotes the channel autocorrelation matrix. $\sigma_n^2$ is the variance of noise contained in ${{{\bf{\hat h}}}_{\rm{p}}^{\rm f}}$ and ${\bf I}$ is a identity matrix. Notice that ${{\bf{W}}_{\rm{d\_MMSE}}}$ does not change w.r.t. the group index $k$. Since we assume a WSSUS channel and the same data structure for all the groups,  the second order channel statistics in (\ref{equ.MMSE}) are invariant w.r.t. $k$ \cite{4202196}. Thus, ${{\bf{W}}_{\rm{d\_MMSE}}}$ is the same for all the groups.

As can be seen from (\ref{equ.MMSE}), MMSE interpolation requires the knowledge of channel correlation and noise variance. The precise channel correlation matrices are cumbersome to be acquired. Though (\ref{equ.MMSE}) is derived under an ideal channel model described in (\ref{equ:chap2.28}), the practical systems may have many unknown imperfections, such as symbol timing offset (STO), carrier frequency offset (CFO), and non-linear distortion. The ideal channel model does not match the real channel environment. As a result, MMSE interpolation in (\ref{equ.MMSE}) usually suffers from performance degradation in real applications.

When STO and CFO are present, the received signal in the time domain will be
\begin{equation}
\label{equ.STOmodel}
\tilde y^{\rm{t}}_n = y^{\rm{t}}_{n-\theta}{e^{ - \frac{{j2\pi n\varepsilon }}{N_{\rm{DFT}}}}},
\end{equation}
where $y^{\rm{t}}_{n-\theta}$ denotes the $\left({n-\theta}\right) \rm{th}$ element of ${\bf{y}}^{\rm{t}}$ and $\theta$ is the timing error. The value of $\theta$ is usually negative and small. $\varepsilon$ is the frequency offset times symbol duration. ${N_{\rm{DFT}}}$ is the size of DFT.

STO causes linear phase shifts among frequency channel responses and thus influences the channel correlation \cite{AthaudageEnhance}. As a result, the channel correlation in (\ref{equ.MMSE}) is inaccurate and should be modified with the knowledge of the timing error. However, the value of the timing error is unknown and a certain loss will result with STO assumed to be 0. CFO causes inter-carrier interference (ICI) and reduces the performance of channel estimation \cite{851324}. Unlike STO, CFO cannot be addressed by properly designing the parameters of the MMSE interpolation even if the value of CFO is known.

To reduce the peak-to-average power ratio (PAPR), the clipping and filtering approach is usually adopted. After clipping, non-linear distortion on the transmitted signal arises, and the distorted signal is given by \cite{8052521}
\begin{equation}
\begin{aligned}
\tilde x^{\rm{t}}_n = \left\{ \begin{gathered}
x^{\rm{t}}_n,\;\;\;\; \left\vert { x^{\rm{t}}_n }\right\vert \le A, \hfill \\
Ae^{j \Phi_n},\; \rm{otherwise}, \hfill \\ 
\end{gathered} \right.
\end{aligned}
\label{equ.Nonlinear}
\end{equation}
where $x^{\rm{t}}_n$ is the $n\rm{th}$ element of ${\bf{x}}^{\rm{t}}$. $A$ represents some amplitude threshold and $\Phi_n$ is the phase of $x^{\rm{t}}_n$.

It is difficult to get the expression of MMSE interpolation under a non-linear model, as shown in (\ref{equ.Nonlinear}). By treating the non-linear distortion as noise, distortion-aware linear MMSE (DA-LMMSE) estimation proposed in \cite{8933050} can improve the estimation performance. However, the required effective noise variance incorporating non-linear distortion is hard to obtain in practical systems.

\subsection{ML-based Channel Estimation}
\label{sec.MLCE}

It is cumbersome to deal with the aforementioned imperfections in MMSE interpolation. Thus, ML-based channel estimation techniques are adopted to address practical imperfections \cite{8052521,8933050}. In ML-based channel estimation, the inputs are the LS estimates $ {{{\bf{\hat h}}}_{{\rm p}\_ k}^{\rm f}} $ and the neural network is a black box to process on the LS estimates and get the outputs ${{\bf{\hat h}}_{{\rm d }\_k}^{\rm{f}}}$, i.e., the estimated channel responses at data subcarriers. Through training, the ML-based estimator can be optimized for a particular hardware configuration and channel. Therefore, information about the channel or the derivation of the channel estimator in closed form is not required and practical imperfections are addressed by exploiting the underlying structures of channel state information during the training process \cite{8663988}.

In ML-based channel estimation, it is important that the channel conditions, especially the practical imperfections to be addressed, in the training procedure and the deployment procedure remain the same. However, most of the existing ML-based channel estimation techniques employ an offline training manner and it is challenging to create high-quality datasets whose features agree well with real-world applications. In addition, the estimator cannot be trained again during transmission even if the channel changes. The ML-based channel estimation with offline training is not amenable to systems where channel conditions change rapidly. It motivates us to explore an online training scheme, where training data is collected during the real-time transmission and the learning-based estimator is trained in real-time to adapt to changing channel conditions.

\section{Linear ML-based Channel Estimation}
\label{sec.LMLCE}

To design an online training scheme, we need to address two problems: how to reduce the amount of required training data and how to collect training data online.

\subsection{Network Structure}

\begin{figure}
\begin{centering}
\includegraphics[scale=0.8]{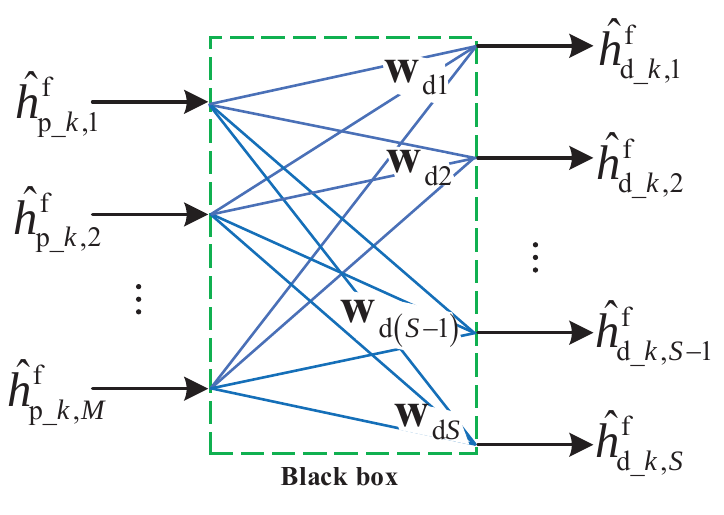}

\end{centering}
\caption{Sketch diagram for LML-based channel estimation.}
\label{Fig.LearningF}
\end{figure}

The size of the training dataset is usually proportional to the number of parameters to be trained in a neural network \cite{8949757}. A deep neural network often has a large number of parameters and thus requires a large dataset for training. For example, if a fully connected neural network with $L$ layers are used and each layer has $U_l$ neurons, there will be $ \sum_{l=1}^{L-1} U_lU_{l+1}$ parameters to be trained. For the sake of reducing the required training data, we use a simple network structure as shown in Fig. \ref{Fig.LearningF}, where there are only $MS$ parameters to be trained. Moreover, only one such network has to be trained since the same estimator is used for all the subcarrier groups in an OFDM symbol, as mentioned in Section \ref{Sec.model}.

In this network, the outputs are directly connected with the inputs through a complex weight matrix ${{\bf{W}}_{\rm{d}}}$, where ${{\bf{W}}_{\rm{d}}} = \left[ {{\bf{w}}_{{\rm d}1}^{\rm T},\ldots ,{\bf{w}}_{{\rm d}S}^{\rm T}} \right]^{\rm T}$. ${\bf{w}}_{{\rm d}s}$ denotes the weight vector connecting the $s\rm{th}$ output ${{{\hat h}}_{{\rm d}\_ k,s}^{\rm f}}$ with the inputs ${{{\bf{\hat h}}}_{\rm{p}}^{\rm{f}}}$, i.e.,
\begin{equation}
\label{equ.LML}
{{\hat h}_{{\rm d}\_k,s}^{\rm{f}}} = {\bf{w}}_{{\rm d}s}^{\rm T} {{{\bf{\hat h}}}_{{\rm p}\_ k}^{\rm f}}.
\end{equation}
We can see that the learning module keeps the same structure as conventional channel estimation methods described in Section \ref{Sec.ConvMeth}, which makes it compatible to the existing implementation. It belongs to the linear channel estimation and thus we call it linear machine learning (LML)-based channel estimation.

We restrict the learning module to a smaller class of functions, i.e., the linear functions, so that the required training data can be reduced greatly. This strategy, which restricts the estimator to a certain class of functions and then finds the best estimator in that class, is also used in \cite{8272484}. Furthermore, the linear learning module is chosen by comparing it with machine learning techniques that are applied for channel estimation in recent literature, including DNN and extreme learning machine (ELM). DNN is a typical deep learning technique, while ELM has a very simple network structure and is the most promising technique to reduce the required training data. Also, it keeps the non-linear fitting ability. The simulation results presented in Fig. \ref{fig3} in Section \ref{Sec.Sim} show that for the proposed online training scheme the linear learning module has better performance than the other two machine learning techniques although those two techniques have better potential performance. It indicates that the training data provided in the proposed online training scheme is not sufficient for those two techniques. In addition, the computational complexity of the learning module is low and its training is fast. The detailed analysis is presented in Subsection \ref{Sec.CompAnal}.

Although the proposed channel estimation method has the same structure as the conventional linear channel estimation methods, the way to obtain the interpolation matrix in (\ref{equ.Int}) is quite different, which is in a data-driven manner. This results in two main advantages of the proposed method. First, under complicated channel models, e.g., non-linear one shown in (\ref{equ.Nonlinear}), the estimator in the proposed method can be directly optimized through training, while the conventional method need the exact expression of the estimator, which is cumbersome to derive. Moreover, the proposed method can adapt to some unknown imperfections. In contrast, the conventional methods usually suffer from performance degradation if those imperfections are not modeled and addressed.

It should be noted that the proposed method loses non-linearities compared with the Deep learning-based channel estimation. As a result, the proposed method may fail to address some severe non-linear imperfections. A DNN may be applied to replace the linear learning module to enable non-linear fitting ability. However, the online training scheme needs to be reconsidered and a much larger training dataset should be provided. We left this problem for our future work.

\subsection{Train the Estimator}
 
To learn the weight matrix ${{\bf{W}}_{\rm{d}}}$, a training dataset is required. Assume the size of the dataset is $T$ and $T>M$, where $M$ is the input dimension of the estimator as shown in Fig. \ref{Fig.LearningF}. The dataset $\cal T$ can be expressed as ${\cal T} =\left\{ {\left( {{\bf x}_{\rm I}\left( 1 \right),{\bf y}_{\rm O}\left( 1 \right)} \right),...,\left({{\bf x}_{\rm I}\left( T \right),{\bf y}_{\rm O}\left( T \right)} \right)} \right\}$, where ${\bf x}_{\rm I}$ and ${\bf y}_{\rm O}$ denote the input and the label for the input, respectively. The issue about how to generate the inputs and labels online is interpreted in the next section.
 
The training is to derive the weight matrix ${{\bf{W}}_{{\rm{d}}*}}$ that minimizes the loss function ${\cal L}_2 = \frac{1}{S}{\left\| {{{\bf{W}}_{{\rm{d}}}}{\bf x}_{\rm I} - {\bf y}_{\rm O}} \right\|_2^2}$ over the dataset ${\cal T}$, i.e., ${{\bf{W}}_{{\rm{d}}*}} = \mathop {\arg }\limits_{{{\bf{W}}_{{\rm{d}}}}} \min \sum\limits_t {\left\| {{{\bf{W}}_{{\rm{d}}}}{\bf x}_{\rm I}\left( t \right) - {\bf y}_{\rm O}\left( t \right)} \right\|_2^2}$. This optimization problem has an analytical solution, which is given by
\begin{equation}
\label{equ.IntLearn}
{{\bf{W}}_{{\rm{d}}*}} ={\bf Y}_{\rm O}{\left( {{\bf X}_{\rm I}} \right)^{\dagger}},
\end{equation}
where ${\bf Y}_{\rm O} = \left[ {{\bf y}_{\rm O}\left( 1 \right),...,{\bf y}_{\rm O}\left( T \right)} \right]$ is a $S \times T$ matrix containing the labels and ${{\bf X}_{\rm I}} = \left[ {{\bf x}_{\rm I}\left( 1 \right),...,{\bf x}_{\rm I}\left( T \right)} \right]$ is a $M \times T$ matrix containing the inputs. The MP generalized inverse ${\left( {{\bf X}_{\rm I}} \right)^{\dagger}}$ can be calculated using the singular value decomposition (SVD) \cite{4012031}.

With the learned weight matrix ${{\bf{W}}_{{\rm{d}}*}}$, channel estimation can be performed to obtain the channel responses at data subcarriers using (\ref{equ.LML}). ${{\bf{W}}_{{\rm{d}}*}}$ is obtained during the transmission of OFDM symbols. This is real-time training and thus we call the training process as online training.

\subsection{ Complexity Analysis }
\label{Sec.CompAnal}
We use the number of complex multiplications (CMs) to measure the computational complexity.
For simplicity, the numbers of CMs for calculating the $M \times T$ ($M<T$) matrix pseudo-inverse and the $M \times M$ matrix inverse are assumed to be $C_{\rm pinv}\left({TM^2} \right)$ and $C_{\rm inv}\left({M^3} \right)$, respectively. We compare the proposed method with the MMSE estimation and the ML-based estimation counterpart, i.e., DNN-based estimation \cite{8052521} and complex ELM (C-ELM) based estimation \cite{8715649}. For a fair comparison, we assume that these methods have the same input dimension $M$ and output dimension $S$ except DNN-based estimation. For DNN-based estimation, we split a complex number into a real part and an
imaginary part. Thus, the input dimension and the output dimension of DNN-based estimation are $2M$ and $2S$, respectively. In DNN-based estimation, there are real-valued multiplications. When calculating the computational complexity, we consider that 4 real-valued multiplications are equivalent to one CM.

The ML-based estimation has two phases, i.e., the training phase and the estimation phase, and we analyze the computational complexity for the two phases individually. The required numbers of CMs for the estimation phase are summarized in Table \ref{tab.CC}, where $N_l$ and $c_l$ denote the number of layers and the number of neurons at the $l\rm{th}$ layer, respectively. $L$ is the number of hidden neurons in the C-ELM. As can be seen, the proposed method has a significantly lower computational complexity compared to the other methods, especially the DNN-based estimation and MMSE estimation.

As for the training phase, the proposed method requires $C_{\rm pinv}\left({TM^2} \right)+T\left({M + S} \right)$ CMs. In C-ELM-based estimation, the calculation of the output weights needs $C_{\rm pinv}\left({TL^2} \right)+T\left({L + S} \right)$ CMs. Since the number of hidden neurons $L$ is usually larger than the input dimension $M$, the number of CMs in C-ELM-based estimation has already exceeded the proposed method when only the calculation of the output weights is taken into account. The training complexity of DNN-based estimation cannot be measured using the number of CMs because it is trained iteratively with back propagation (BP).  The time consumption of training is hard to satisfy the latency constraint in practical uses and thus DNN-based estimation is trained in an offline manner \cite{8052521}. ELM is  well known for its fast training capability. The training complexity of the proposed method is even lower than C-ELM. Hence, the training of the proposed method can be performed online.

\begin{table}  
\caption{Complexity Comparison}
\label{tab.CC}  
\begin{center}  
\begin{tabular}{|c|c|}  
\hline  
Estimators & Numbers of CMs  \\
\hline
\hline
C-ELM-based \cite{8715649}& $MSL$   \\  
\hline   
DNN-based \cite{8052521}& $MS\sum_{l=1}^{N_l}c_lc_{l-1}$    \\
\hline
MMSE & $C_{\rm inv}\left({M^3} \right)+M\left({M + S}\right)$    \\  
\hline
Proposed  & $MS$   \\  
\hline
\end{tabular}
\end{center}    
\end{table}  

\section{ Generation of Training Data}
\label{sec.Generation}

The training data $ \left( {{\bf x}_{\rm I}\left( t \right),{\bf y}_{\rm O}\left( t \right)} \right) $ given above is not described in detail, where $t$ is the serial index in the dataset. In this section, we first introduce a novel training data structure, which can be collected online. Then, we present the analysis of the feasibility of the novel training data structure. Based on this training data structure, we propose two alternative training data generation schemes.

\subsection{ Training Data Structure }
Conventionally, the input $ {\bf x}_{\rm I}$ is the LS estimates at pilot subcarriers, and the label $ {\bf y}_{\rm O}$ is the ideal output, i.e., the actual values of channel responses at the corresponding data subcarriers. The input can be obtained by transmitting pilot signals, while the label is hard to provide during the transmission of OFDM symbols. However, we find that the LS estimates at data subcarriers can replace their actual values to be the labels, whose feasibility analysis is presented in the next subsection. This label can be obtained by transmitting extra pilot signals (the signals at data subcarriers are also known at the receiver) or in a decision-directed way, which is illustrated in Section \ref{sec.TDGS}. With this structure, training data can be collected during the transmission of OFDM symbols.

The input-output pair for training data is given by ${\bf x}_{\rm I}\left( t \right)={{{\bf{\hat h}}}_{{\rm p}\_ t}^{\rm f}}$ and ${\bf y}_{\rm O}\left( t \right)={\hat {\bf{ h}}_{{\rm d}\_t}^{\rm{f}}}$. ${{{\bf{\hat h}}}_{{\rm p}\_ t}^{\rm f}}$ is similar to ${{{\bf{\hat h}}}_{{\rm p}\_ k}^{\rm f}}$ in (\ref{equ.Int}), which contains the LS estimates at pilot subcarriers. ${\hat {\bf{ h}}_{{\rm d}\_t}^{\rm{f}}}$ contains the LS estimates at corresponding data subcarriers.

This structure makes it possible that the collected training data and the transmitted signals to be recovered are from the same OFDM frame. This is quite meaningful for a machine learning-based channel estimation technique since it guarantees that the channel conditions in the training phase and the deployment phase are the same. The training data may capture practical imperfections, such as non-linear distortion, that cannot be well described by tractable models. As a result, the impact on the estimation performance of these imperfections may be alleviated through training over the collected data.

\subsection{ Feasibility Analysis}

It has been demonstrated in the literature that an estimator can be learned through training on a dataset labeled by true channel responses, i.e., $\left\{ {...,\left( {{{\bf{\hat h}}}_{{\rm p}\_t}^{\rm{f}},{{{\bf{ h}}}^{\rm{f}}_{{\rm d}\_t}}} \right),...} \right\}$, where ${{{\bf{ h}}}^{\rm{f}}_{{\rm d}\_t}}$ denotes the vector containing true channel responses at data subcarriers. However, in the proposed method, ${{{\bf{ h}}}^{\rm{f}}_{{\rm d}\_t}}$ in the dataset is replaced by its estimation result ${{\hat {\bf{ h}}}^{\rm{f}}_{{\rm d}\_t}}$. The labels of training data ${{\hat {\bf{ h}}}^{\rm{f}}_{{\rm d}\_t}}$ are not the ideal outputs ${{{\bf{ h}}}^{\rm{f}}_{{\rm d}\_t}}$ and influenced by noise. Furthermore, the performance of labels is usually not satisfactory since we use a simple method, i.e., LS estimation, to obtain the labels. Therefore, we demonstrate the feasibility of the proposed training data in this subsection.

First, it should be noticed that the training of the estimator is not meant to achieve the performance of the labels, i.e., the LS estimates. This is because the learning module does not exactly produce the labels after training. The performance of the learned estimator is usually not the same as that of labels. This is verified in simulation experiments and the simulation results are given in Fig. \ref{fig.MSE}.

Next, we show that like the conventional training data, the proposed training data has the property that the optimal estimator can be learned when the average loss function approaches its expectation. This condition usually holds when the size of training data gets infinitely large. When true channel responses are used as labels, the loss function is ${\cal L}_2 = \frac{1}{S}{\left\| {{{{\hat {\bf h}}_{{\rm dML}\_t}}^{\rm f}} - {{{\bf{ h}}}_{{\rm d}\_t}^{\rm f}}} \right\|_2^2}$, where ${{{\hat {\bf h}}_{{\rm dML}\_t}}^{\rm f}}={{\bf{W}}_{\rm{d}}} {{{\bf{\hat h}}}_{{\rm p}\_ t}^{\rm f}}$ denotes the output of the estimator. Its expectation is the mean square error (MSE). The average loss function can be minimized through training and thus the learned estimator has the minimum MSE since the average loss function is the MSE.

When LS estimates are used as labels, the loss function is ${\cal L}_2 = \frac{1}{S}{\left\| {{{{\hat {\bf h}}_{{\rm dML}\_t}}^{\rm f}} - {{\hat {\bf{ h}}}_{{\rm d}\_t}^{\rm f}}} \right\|_2^2}$. Then, the expectation of the loss function becomes $\sigma _{{\text{MSE}}}^2 + \sigma _{\rm{LS}}^2$ as derived in Lemma 1, where $\sigma _{{\text{MSE}}}^2$ is the MSE of ${{\hat {\bf h}}_{{{\rm d ML}}\_t}}$ and $\sigma _{\rm{LS}}^2$ is the MSE of LS estimation. During training, the parameters of the estimator are optimized by minimizing the average loss function, which can be formulated as $\mathop {\arg }\limits_{{{\bf{W}}_{\rm{d}}}} \min \left( {\sigma _{{\rm{MSE}}}^2 + \sigma _{{\rm{LS}}}^2} \right)$. Since $\sigma _{\rm{LS}}^2$ is not influenced by ${{{\bf{W}}_{\rm{d}}}}$, we have 
$$\mathop {\arg }\limits_{{{\bf{W}}_{\rm{d}}}} \min \left( {\sigma _{{\rm{MSE}}}^2 + \sigma _{{\rm{LS}}}^2} \right) = \mathop {\arg }\limits_{{{\bf{W}}_{\rm{d}}}} \min \sigma _{{\rm{MSE}}}^2.$$
It indicates that the learned estimator also has the minimum MSE. Therefore, it is proved that when the average loss function approaches its expectation, using the proposed training data can learn the optimal estimator like the conventional training data with ideal labels.

Lemma 1: $\mathbb{E}\left[{ \frac{1}{S}{\left\| {{{{\hat {\bf h}}_{{\rm dML}\_t}}^{\rm f}} - {{\hat{\bf{ h}}}_{{\rm d}\_t}^{\rm f}}} \right\|_2^2} }\right] = \sigma _{{\rm{MSE}}}^2 + \sigma _{{\rm{LS}}}^2$.

Proof: 
$\mathbb{E}\left[ { \frac{1}{S}{\left\| {{{{{\hat {\bf h}}_{{\rm dML}\_t}}}^{\rm f}} - {{\hat {\bf{ h}}}_{{\rm d}\_t}^{\rm f}}} \right\|_2^2} } \right]$ can be expressed as  
\begin{equation}
\label{equ.ssee}
\begin{gathered}
  \mathbb{E}\left[ {{{\left|  {{{{\hat h}_{{\rm dML}\_t,s}}^{\rm f}} - {{\hat h}_{{\rm d}\_t,s}^{\rm f}}}  \right|}^2}} \right] \hfill \\
   = \mathbb{E}\left[ {\left( {{{{\hat h}_{{\rm dML}\_t,s}}^{\rm f}} - {{\hat h}_{{\rm d}\_t,s}^{\rm f}}} \right){{\left( {{{{\hat h}_{{\rm dML}\_t,s}}^{\rm f}} - {{\hat h}_{{\rm d}\_t,s}^{\rm f}}} \right)}^ * }} \right] \hfill \\
   = \mathbb{E}\left[ {\left( {{{{\hat h}_{{\rm dML}\_t,s}}^{\rm f}} - {{ h}_{{\rm d}\_t,s}^{\rm f}}-n_s} \right)} \right. \hfill \\
  \;\;\;\;\;\;\;\; \cdot \left. {{{\left( {{{{\hat h}_{{\rm dML}\_t,s}}^{\rm f}} - {{ h}_{{\rm d}\_t,s}^{\rm f}}-n_s} \right)}^ * }} \right], \hfill \\ 
\end{gathered}  
\end{equation}
where $n_s$ denotes the estimation noise in ${{\hat h}_{{\rm d}\_t,s}^{\rm f}}$. $n_s$ is independent of ${{{\bf{\hat h}}}_{{\rm p}\_t}^{\rm{f}}}$, and thus is independent of ${{{\hat h}_{{\rm dML}\_t,s}}^{\rm f}}$ since ${{{\hat h}_{{\rm dML}\_t,s}}^{\rm f}}$ is the linear combination of ${{{\bf{\hat h}}}_{{\rm p}\_t}^{\rm{f}}}$, i.e., ${{{\hat h}_{{\rm dML}\_t,s}}^{\rm f}} = {\bf{w}}_{{\rm d}s}{{{\bf{\hat h}}}_{{\rm p}\_t}^{\rm{f}}}$. (\ref{equ.ssee}) can be simplified as
\begin{equation}
\begin{gathered}
  \mathbb{E}\left[ {{{{\left|  {{{{\hat h}_{{\rm dML}\_t,s}}^{\rm f}} - {{\hat h}_{{\rm d}\_t,s}^{\rm f}}}  \right|}^2}}} \right] \hfill \\
   = \mathbb{E}\left[ {\left( {{{{\hat h}_{{\rm dML}\_t,s}}^{\rm f}} - {{ h}_{{\rm d}\_t,s}^{\rm f}} }\right){{\left( {{{{\hat h}_{{\rm dML}\_t,s}}^{\rm f}} - {{ h}_{{\rm d}\_t,s}^{\rm f}}} \right)}^ * }} \right] \hfill \\
  \;\;\;\; + \mathbb{E}\left[ {{n_s}n_s^ * } \right] \hfill \\
   = \sigma _{{\text{MSE}}}^2 + \sigma _{\rm{LS}}^2 . \hfill \\ 
\end{gathered}  
\end{equation}

The above analysis result can be intuitively explained as follows. When the learning module is trained using one labeled sample, the noise in the label may cause the parameters in the learning module to be modified in the wrong direction. However, the influence of noise may vanish with the accumulation of training data. This is because the noise in labels, i.e., the LS estimates, is zero-mean Gaussian noise \cite{4267831} and the errors in modifying the parameters caused by noise may cancel each other out.

In fact, label noise has been investigated in the machine learning field and it is shown that symmetric label noise may not affect the training performance \cite{6685834}. In our proposed training data, the noise in labels is also symmetric since it is zero-mean Gaussian noise. Furthermore, in addition to the above analysis under the ideal condition, we also conduct simulation experiments to demonstrate the feasibility of our proposed training data in practical situations, and the results and the corresponding analysis are given in Section \ref{Sec.Sim}. In Fig. \ref{fig.DataNum}, the performance of the learned estimator using the proposed training data is compared with that using true channel responses. From the results, it is shown that when the size of training data is sufficiently large, which can be achieved in practice, the performance degradation caused by the noise in labels is quite small.

\subsection{ Training Data Generation Schemes }
\label{sec.TDGS}

In this section, we give two alternative training data generation schemes.

\textbf{\emph{In a pilot-aided manner}}: An intuitive approach is to transmit block pilot symbols, in which all the subcarriers convey pilot signals, ahead of conventional data symbols so that the channel responses at "data" subcarriers can also be estimated using LS estimation. We call this approach as pilot aided training data generation (PATDG) scheme.

Training data is generated based on the LS estimation results using the block pilot symbol, i.e.,
\begin{equation}
\label{equ.BlPiLS}
{\hat {\bf{h}}^{\rm{f}}} = \left( {{\bf{X}}^{\rm{f}}} \right) ^{-1}{\bf{y}}^{\rm{f}}.
\end{equation}

\begin{figure}[htb]
\begin{centering}
\includegraphics[scale=0.65]{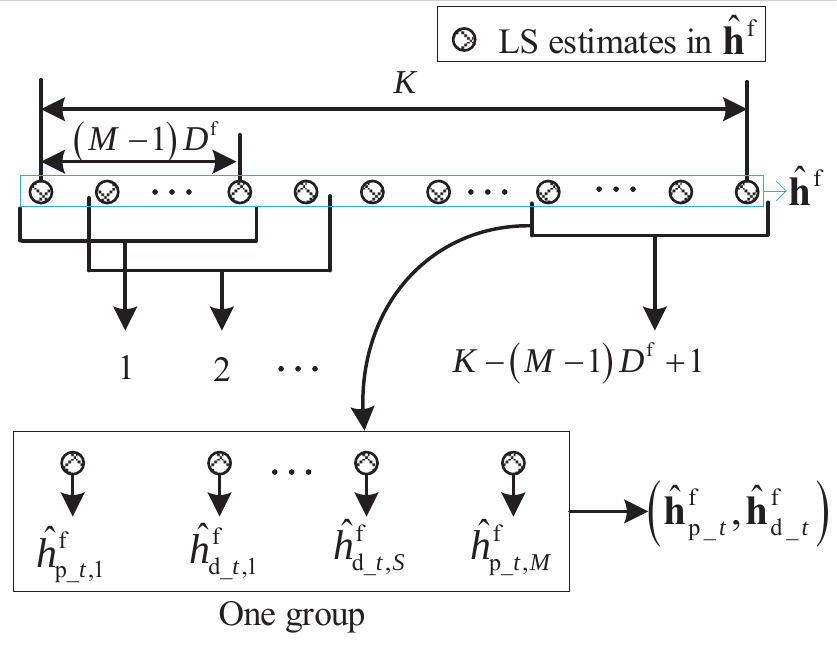}

\end{centering}
\caption{Sketch diagram for illustrating how to generate sample pairs.}
\label{Fig.GIP}
\end{figure}

Considering the transmission efficiency, block pilot symbols should be as few as possible. Therefore, these pilot symbols should be fully utilized to generate training data. Fig. \ref{Fig.GIP} illustrates the proposed training data generation scheme. The consecutive $(M-1){D^{\rm{f}}}$ LS estimates form one group, in which $M$ items work as the estimates of pilot subcarriers ${{{\bf{\hat h}}}_{{\rm p}\_t}^{\rm{f}}}$ and $S$ items work as the estimates of data subcarriers ${{{\bf{\hat h}}}_{{\rm d}\_t}^{\rm{f}}}$. One such group thus can provide one pair of training data $\left({{{{\bf{\hat h}}}_{{\rm p}\_t}^{\rm{f}}},{{{\bf{\hat h}}}_{{\rm d}\_t}^{\rm{f}}}} \right)$. At most $K - (M-1){D^{\rm{f}}}+1$ such groups can be generated based on ${\hat {\bf{h}}^{\rm{f}}}$, as shown in Fig. \ref{Fig.GIP}. In this way, $K - (M-1){D^{\rm{f}}}+1$ sample pairs can be obtained based on ${\hat {\bf{h}}^{\rm{f}}}$. If $N_{\rm p}$ block pilot symbols are transmitted in the training phase, $N_{\rm p}(K - (M-1){D^{\rm{f}}}+1)$ sample pairs can be provided with the proposed training data generation scheme. The dataset $\cal T$ is given by ${\cal T} =\left\{ {\left( {{{{\bf{\hat h}}}_{{\rm p}\_1}^{\rm{f}}},{{{\bf{\hat h}}}_{{\rm d}\_1}^{\rm{f}}}} \right),...,\left({{{{\bf{\hat h}}}_{{\rm p}\_T}^{\rm{f}}},{{{\bf{\hat h}}}_{{\rm d}\_T}^{\rm{f}}}} \right)} \right\}$, and the size of ${\cal T}$ is 
\begin{equation}
T = N_{\rm p}(K - (M-1){D^{\rm{f}}}+1).
\label{equ.SamNum}
\end{equation}

Note that the modulation mode of block pilot symbols should be the same with data symbols.

\begin{figure}[htb]
\begin{centering}
\includegraphics[scale=0.8]{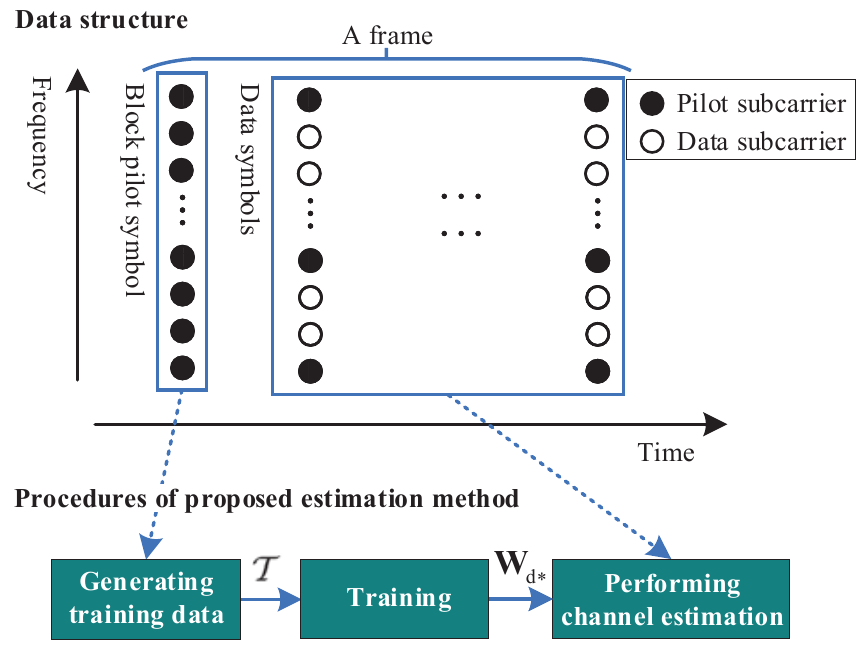}

\end{centering}
\caption{Data structure and procedures of the proposed channel estimation method with PATDG.}
\label{DDDSCP}
\end{figure}

Under certain conditions, one block pilot symbol is sufficient, i.e., $N_{\rm p}=1$. This will be demonstrated by simulation results in Section \ref{Sec.Sim}. The procedures of the proposed estimation method with PATDG along with the data structure are shown in Fig. \ref{DDDSCP}. When receiving an OFDM frame, the receiver first train the estimator using the training data provided by the block pilot symbol. Then, the trained estimator is used to obtain the channel responses at data subcarriers of OFDM symbols. The estimator is re-trained for each OFDM frame and thus can adapt to quickly changing channel conditions.

\textbf{\emph{In a decision-directed manner}}: Different from the PATDG, we propose another scheme called the decision-directed training data generation (DDTDG) scheme, which does not require extra pilot overhead for the generation of training data. The main idea behind DDTDG is inspired by the decision-directed channel estimation (DDCE), which uses the detected symbols as pilots to track the channel. Similarly, we can collect training data in a decision-directed way.

In the proposed DDTDG scheme, the data structure is the same as the one shown in Fig. \ref{Fig.DivSymbol}. Initial channel estimates are first obtained with a simple channel estimation algorithm, which is the LS method with linear-interpolation. Using initial estimation, the single tap equalization in the frequency domain can be performed followed by the hard decision to recover the transmitted data \cite{8663458}. Then, the recovered data is fed back and treated as a block pilot symbol to calculate ${\hat {\bf{h}}^{\rm{f}}} $ using (\ref{equ.BlPiLS}). In the sequel, training data can be generated with ${\hat {\bf{h}}^{\rm{f}}}$ as the PATDG scheme.

In order to explain the principle of DDTDG specifically, we use the procedure of the proposed channel estimation method with DDTDG to present how it works. The procedure is shown in Algorithm \ref{alg.IterCE} and Steps 1 to 5 represent the procedure of DDTDG. Initial estimation ${{\bf{\hat h}}}_{\rm d}^{\rm{f}}$ is first obtained using linear interpolation. The result ${{\bf{\hat h}}}_{\rm d}^{\rm{f}}$ is then used to recover the transmitted data and generate training data. With the collected training data, the estimator can be learned using (\ref{equ.IntLearn}) and the interpolation matrix ${{\bf{W}}_{\rm{d}}}$ can be updated. With the updated ${{\bf{W}}_{\rm{d}}}$, the output of the algorithm ${{\bf{\hat h}}}_{\rm d}^{\rm{f}}$ can be calculated using (\ref{equ.Int}).

\begin{algorithm}[htbp]
\caption{Proposed channel estimation algorithm with DDTDG }
\label{alg.IterCE}
\centering
\begin{algorithmic}[1]	
\REQUIRE ~~\\
pilot signals ${{\bf{X}}_{\rm{p}}^{\rm f}}$, received signals ${{\bf{y}}^{\rm f}}$;
\ENSURE ~~\\
channel responses at data subcarriers ${{\bf{\hat h}}}_{\rm d}^{\rm{f}}$;
\STATE  calculate ${{{\bf{\hat h}}}_{{\rm p}\_k }^{\rm f}}(k=1,...,G)$ using (\ref{equ.LS});
\STATE  calculate ${{\bf{\hat h}}}_{\rm d}^{\rm{f}}$ using ${{{\bf{\hat h}}}_{{\rm p}\_k }^{\rm f}}$ with linear interpolation;
\STATE  recover the transmitted data with single tap equalization and hard decision;
\STATE calculate ${\hat {\bf{h}}^{\rm{f}}}$ using (\ref{equ.BlPiLS}) with the recovered data regarded as pilots;
\STATE generate training data ${\cal T}$ using ${\hat {\bf{h}}^{\rm{f}}}$ as shown in the PATDG scheme;
\STATE obtain ${{\bf{W}}_{\rm{d}}}$ based on ${\cal T}$ using (\ref{equ.IntLearn});
\STATE update ${{\bf{\hat h}}}_{\rm d}^{\rm{f}}$ using ${{\bf{W}}_{\rm{d}}}$ based on (\ref{equ.Int}).
\end{algorithmic}
\end{algorithm}

It can be seen that the Algorithm 1 does not need the extra pilot overhead and therefore it does not cause any loss in spectral efficiency while generating training data.

\section{ Simulation Results}
\label{Sec.Sim}

We simulate three typical scenarios, including the linear channel with perfect synchronization, the linear channel with imperfect synchronization, and the non-linear channel depicted in (\ref{equ.Nonlinear}) with perfect synchronization. Under scenario 1, there is a closed-form expression for the optimal estimator, i.e., the MMSE channel estimation shown in (\ref{equ.MMSE}), so that we have a good baseline to verify the performance of the proposed method. Under scenario 2 and 3, we examine the adaption of the proposed method to practical imperfections, including STO, CFO, and non-linear distortion. Moreover, we validate the use of a linear learning module for our proposed online training scheme through comparisons with other machine learning techniques, e.g., deep neural networks, under scenario 3. Since deep neural networks have better potential under this scenario due to their stronger fitting abilities, convincing evidence can be provided for the superiority of a linear learning module in the proposed scheme if the linear learning module still outperforms other machine learning techniques.

The simulations are based on the model described in Section \ref{Sec.model}. The system parameters are shown in Table \ref{tab.para} and the channel parameters for the three scenarios are shown in Table \ref{tab.chpara}, where $\theta _{\rm min}$ and $\varepsilon_{\rm max}$ represent the minimum value of STO and the maximum value of normalized CFO, respectively. The STO and the normalized CFO are assumed to be uniformly distributed within $\left[ {\theta _{\rm min},...,0} \right]$ and $ \pm {\varepsilon _{\max }}$, respectively. For the non-linear distortion, the amplitude threshold $A$ in (\ref{equ.Nonlinear}) is set as the root mean square of signal, as in \cite{8052521}. We use normalized MSE (NMSE), as shown in (\ref{equ.NMSE}), to measure the estimation performance. When MSE is hard to be calculated, bit error rate (BER) is used and zero-forcing (ZF) equalization and hard decision are adopted to recover the information bits. The NMSE and BER are obtained by averaging over 5000 independent Monte Carlo runs.
\begin{equation}
\label{equ.NMSE}
{\rm NMSE} = {{{\mathbb E} \left[ {{\left\| {\hat {\bf{h}}_{{\rm{d}}\_k}^{\rm{f}} - {\bf{h}}_{{\rm{d}}\_k}^{\rm{f}}} \right\|_2^2}} \right]} \over {{\mathbb E} \left[ {\left\| {{\bf{h}}_{{\rm{d}}\_k}^{\rm{f}}} \right\|_2^2} \right]} }.
\end{equation}

\begin{table} 
\caption{System Parameters}
\label{tab.para} 
\begin{tabular}{|c|c||c|c|}
\hline 
Number of subcarriers & 512 & CP length & 128 \\ 
\hline 
Symbol period (with CP) & 32 $\mu s$ & Bandwidth & $10\rm{MHz}$ \\
\hline
Channel model \cite{ITUR} & Pedestrian B & Modulation & QPSK \\ 
\hline
Pilot interval & ${D^{\rm{f}}}=3$ & Estimator taps & $M=2$ \\ 
\hline
Block pilot symbols & $N_{\rm p}=1$&Data symbols& $N_{\rm d}=9$\\
\hline
Available subcarriers & $K= 410$& & \\
\hline
\end{tabular} 
\end{table} 

\begin{table} 
\caption{Channel Parameters}
\label{tab.chpara} 
\begin{center} 
\begin{tabular}{|c|c|c|c|} 
\hline 
& $\theta _{\rm min}$ & $\varepsilon_{\rm max}$ & Non-linear distortion\\ 
\hline 
Scenario 1 & 0 & 0 & No \\
\hline
Scenario 2 & -20, -40 & 0.01, 0.05 & No \\ 
\hline
Scenario 3 & 0 & 0 & Yes \\ 
\hline
\end{tabular} 
\end{center} 
\end{table}

\subsection{Simulation Results under Scenario 1}
We demonstrate the feasibility of the proposed training data labeled by estimated channel responses through simulation experiments under this scenario and test the robustness of the proposed channel estimation method.

\begin{figure}[!htb]
\begin{centering}
\includegraphics[scale=0.55]{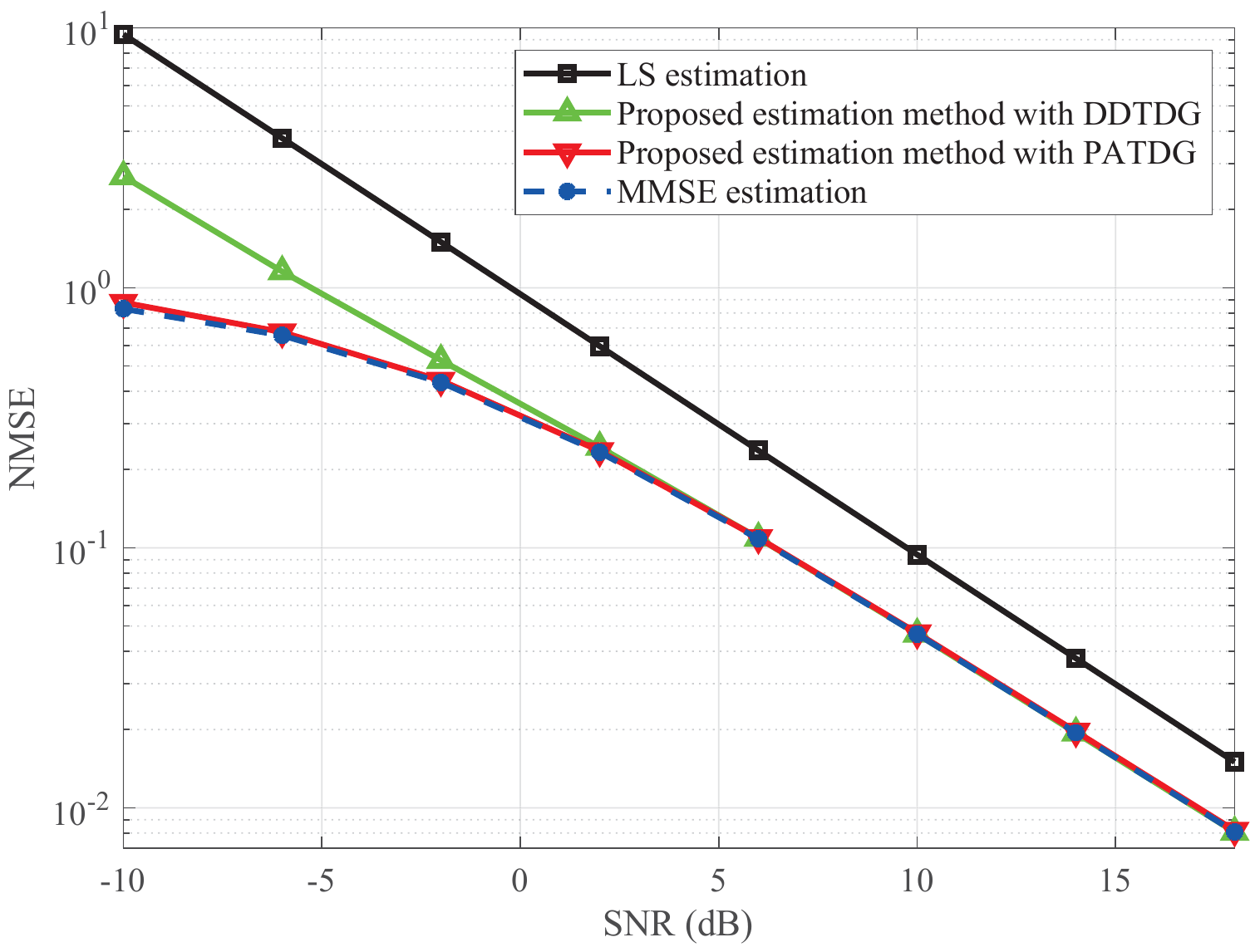}
\par\end{centering}
\caption{The NMSE performance of MMSE estimation, LS estimation, and the proposed method with PATDG and DDTDG vs. SNR under scenario 1.}
\label{fig.MSE}
\end{figure}

Fig. \ref{fig.MSE} compares the NMSE performance of the proposed channel estimation method with the two training data generation schemes. We can see that although labels in the proposed training data structure are the estimated channel responses using LS estimation, the proposed method with PATDG and that with DDTDG both achieve much better performance as opposed to LS estimation. It shows that the performance of learned estimation is not related to the estimation performance of labels. Moreover, the proposed method with PATDG is robust to noise and yields a result close to MMSE estimation for all signal-to-noise ratios (SNRs), while that with DDTDG suffers from significant performance degradation at lower SNRs. This is because DDTDG uses the detected data as the transmitted data to calculate labels, and the wrongly detected data generates bad labels and thus causes performance degradation. This problem becomes more severe at low SNRs, where the ratio of wrongly detected data raises. Therefore, the DDTDG scheme is sensitive to the noise level and not suitable for the systems that operate in low SNR environments. However, for the systems where the power of the received signal is strong and spectrum efficiency is given preference, DDTDG is a better choice than PATDG. In the following simulations, we only use PATDG for the proposed method. This is because the two schemes perform closely at high SNRs and PATDG can represent the proposed training data structure better than DDTDG. The labels in PATDG are calculated using the accurate transmitted data.

\begin{figure}[!htb]
\begin{centering}
\includegraphics[scale=0.475]{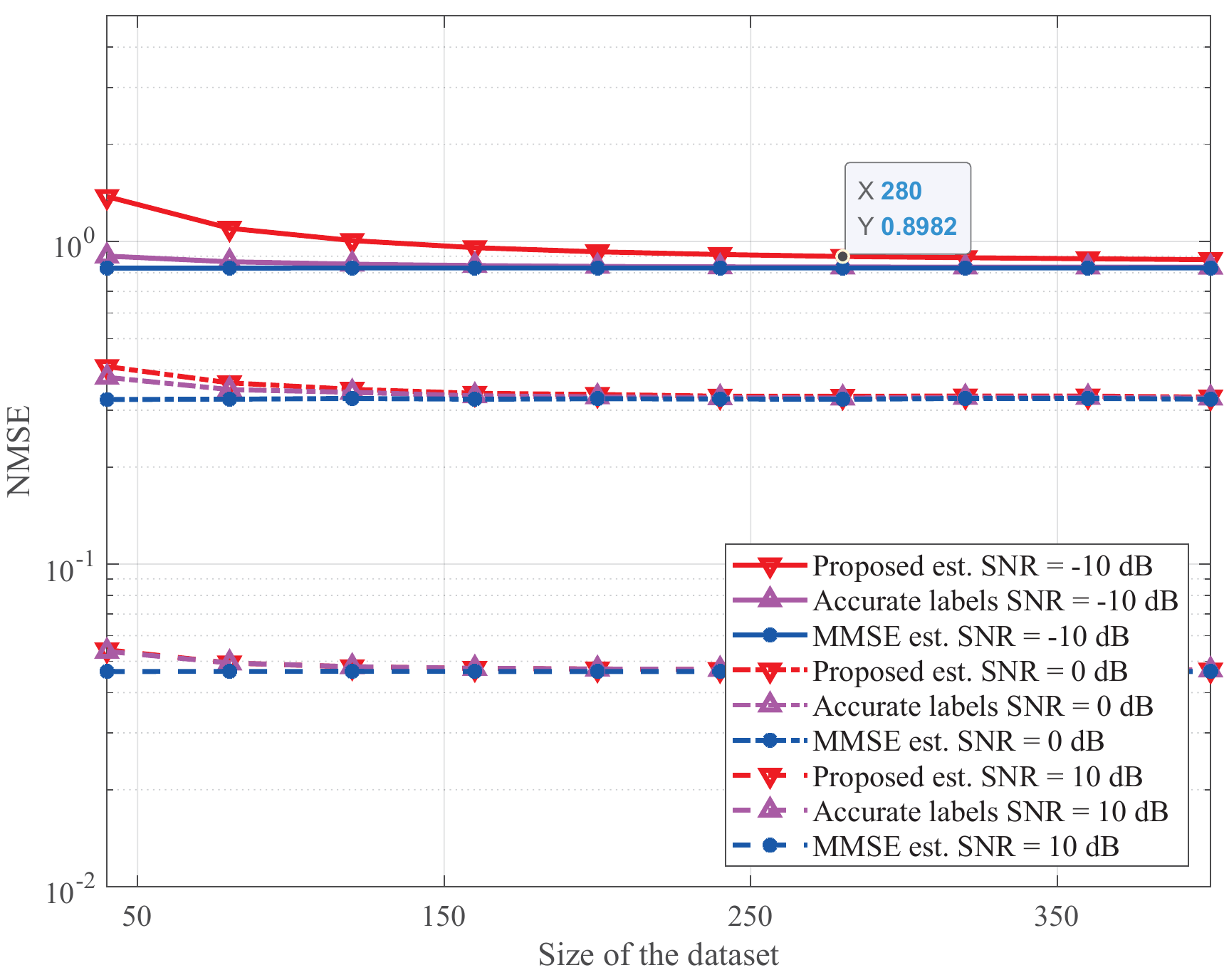}
\par\end{centering}
\caption{The NMSE performance of the proposed method and MMSE estimation vs. the size of the dataset under scenario 1. The proposed method is trained with estimated labels and accurate labels, respectively.}
\label{fig.DataNum}
\end{figure}

To provide further insights on the proposed training data structure, we compare it with the conventional one, i.e., the training data labeled by actual channel responses. The results are presented in Fig. \ref{fig.DataNum}. First, through the comparison between the proposed method and the one with the training data labeled by actual channel responses, we can see that the performance loss caused by the estimation error in labels decreases with the increasing SNR. Moreover, when the size of the dataset is over 280, performance loss becomes quite small even at the SNR of -10 dB and the performance of the proposed method gets close to that of MMSE estimation. According to (\ref{equ.SamNum}), a block pilot symbol can provide $408$ samples in our simulations. Therefore, one symbol is enough, i.e., $N_{\rm p}=1$, and an estimator with near-optimal performance can be learned even at a low SNR.

\begin{figure}[!htb]
\begin{centering}
\includegraphics[scale=0.55]{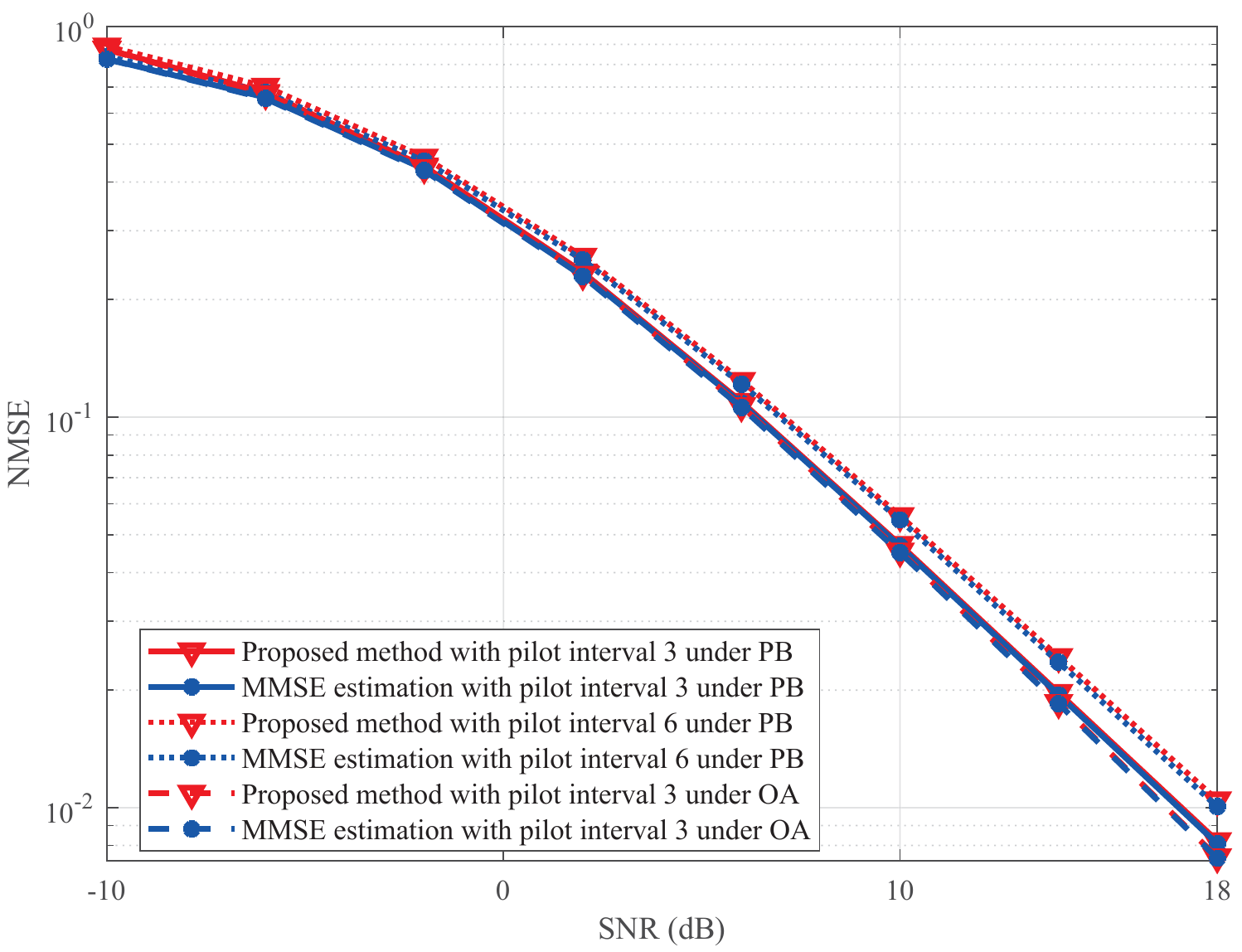}
\par\end{centering}
\caption{The NMSE performance of the proposed method and MMSE estimation vs. SNR for different pilot intervals and channel models under scenario 1.}
\label{fig.PilotDens}
\end{figure}

Fig. \ref{fig.PilotDens} shows the NMSE performance for two pilot intervals ${D^{\rm{f}}}$ and besides Pedestrian B (PB) the performance under an additional channel model, i.e., Office A (OA), is displayed as well. Notice that the results of the proposed method are all close to the MMSE estimation for different pilot intervals and for different channel models. The simulation results demonstrate the robustness of the proposed method to the pilot intervals and propagation environments.

\subsection{Simulation Results under Scenario 2}

Under this scenario, we compare the proposed method with two conventional estimation methods, linear interpolation and a fixed MMSE estimation method proposed in \cite{AthaudageEnhance}. We choose the linear interpolation because it is a widely used method in practical systems due to its simplicity. The fixed MMSE estimation uses the STO statistics to improve the performance and we call it the average MMSE (AMMSE) estimation in this paper.

\begin{figure}
\begin{centering}
\includegraphics[scale=0.475]{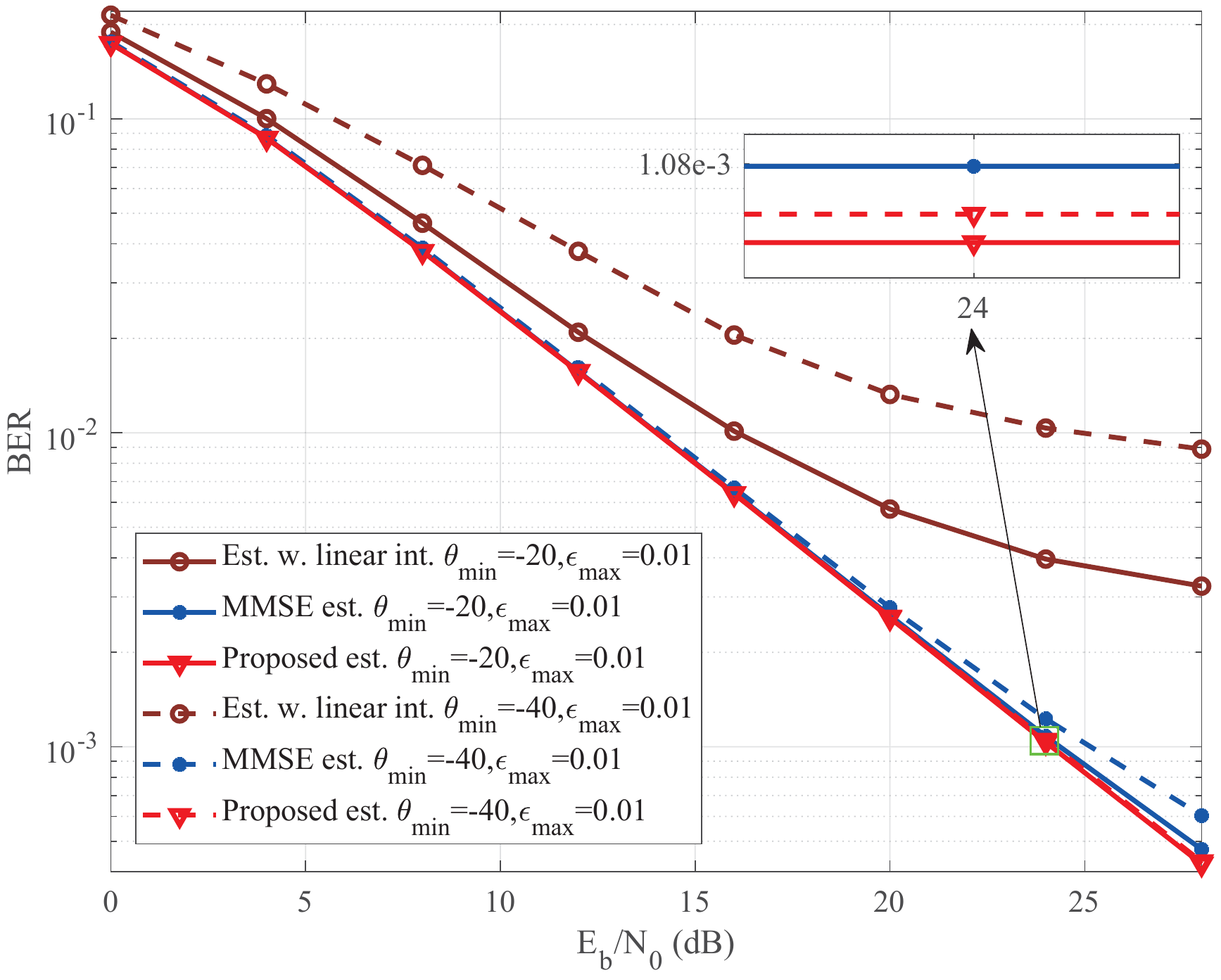}

\end{centering}
\caption{The BER performance of the proposed method, estimation with linear interpolation, and average MMSE estimation vs. $E_b/N_0$ for different $\theta _{\rm min}$ under scenario 2.}
\label{fig2}
\end{figure}

Fig. \ref{fig2} depicts the BER curves under various values of $\theta _{\rm min}$. It can be seen that STO corrupts the linear interpolation severely. In addition, the comparison between the proposed method and the average MMSE estimation, in which the used STO statistics are assumed to be accurate, reveals that the proposed method addresses STO even better. It outperforms the average MMSE estimation especially when $\theta _{\rm min}=-40$ at high SNR regions. It shows that the proposed method can learn the timing error based on training data and then deal with it very well.

\begin{figure}
\begin{centering}
\includegraphics[scale=0.475]{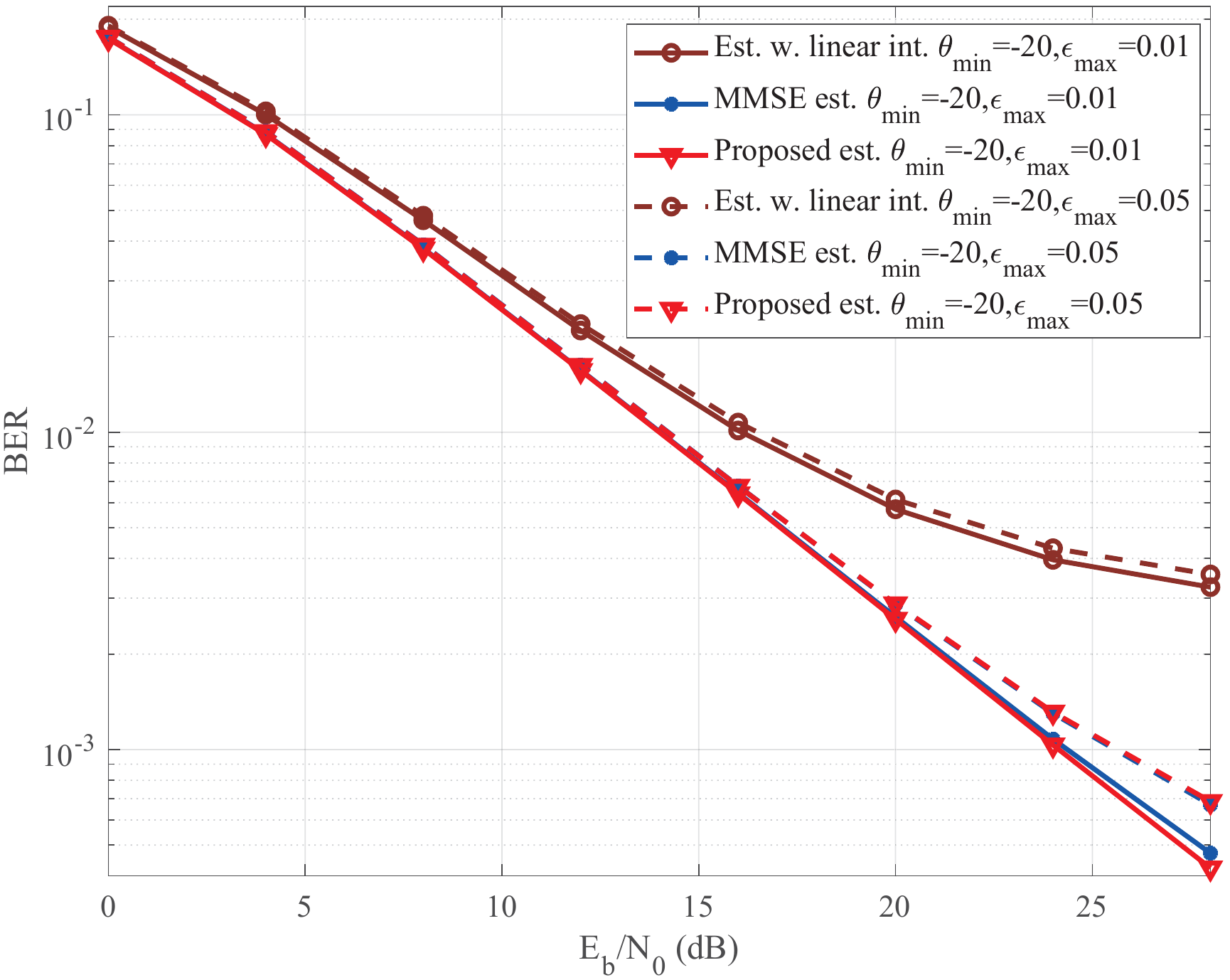}

\end{centering}
\caption{The BER performance of the proposed method, estimation with linear interpolation, and average MMSE estimation vs. $E_b/N_0$ for different $\varepsilon_{\rm max}$ under scenario 2.}
\label{fig.CFO}
\end{figure}

Fig. \ref{fig.CFO} presents the BER curves under different values of $\varepsilon_{\rm max}$. As can be seen, the proposed method and the average MMSE estimation suffer from similar performance degradation when $\varepsilon_{\rm max}$ grows to $0.05$, which shows that the proposed method fails to address CFO as well. This is because the proposed method employs a linear learning module and CFO cannot be addressed through a linear operation of the estimator. It shows that the proposed method is not adaptive to the practical imperfections which cannot be compensated by a linear operation.

\subsection{Simulation Results under Scenario 3}
To validate the linear learning module, we employ the C-ELM \cite{8715649} and the three-layer DNN \cite{8052521} to the black box in Fig. \ref{Fig.LearningF} for benchmarking purposes, respectively. The number of hidden neurons $L$ in C-ELM is set as $8$. The numbers of neurons in each layer of DNN are $4$, $8$, $4$, respectively. The number of neurons per layer is chosen through extensive simulation experiments. Specifically, those chosen parameters for C-ELM and DNN show better or at least similar performance compared with other parameter settings according to our experiment results.

\begin{figure}
\begin{centering}
\includegraphics[scale=0.5]{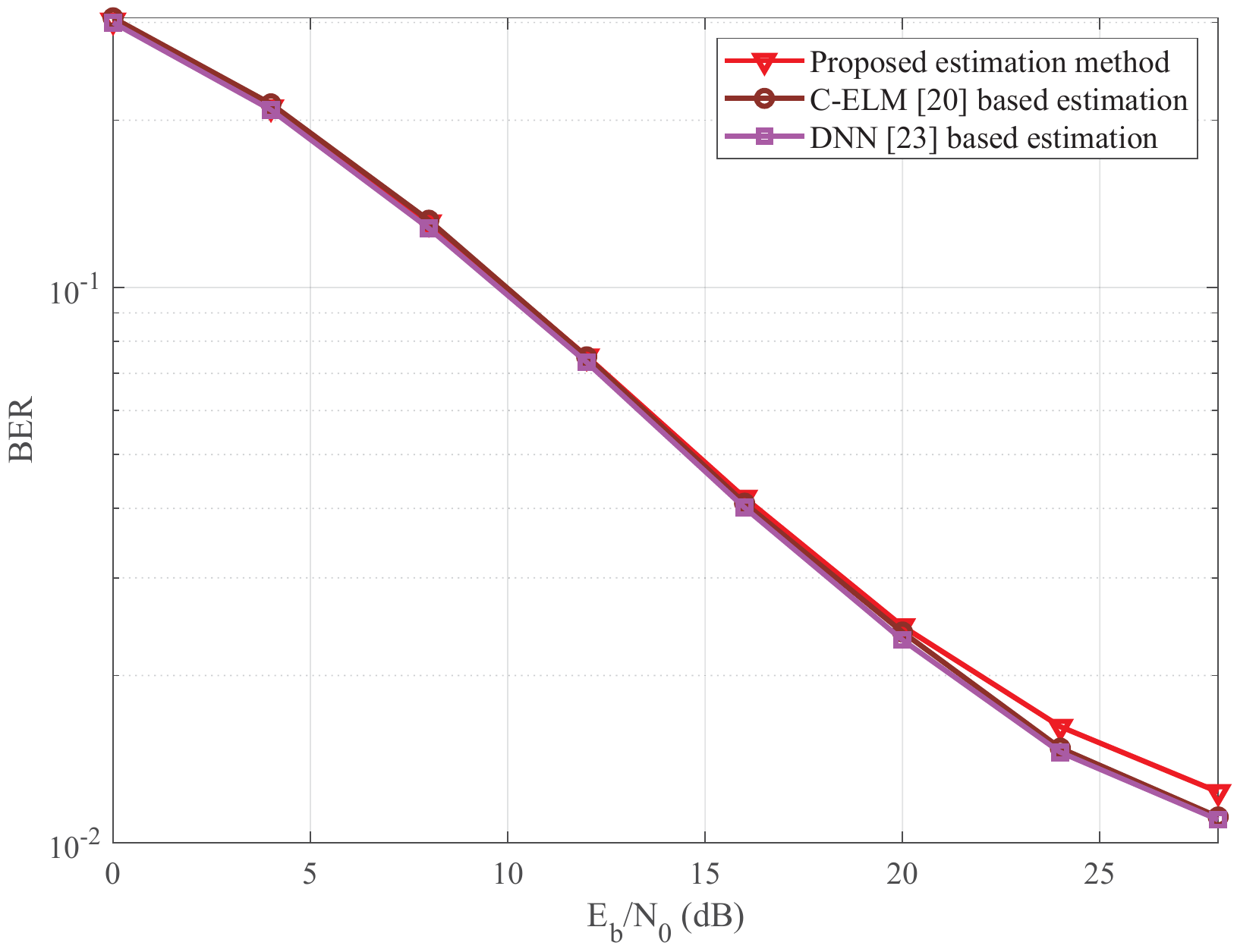}

\end{centering}
\caption{The BER performance of the proposed method, C-ELM-based estimation, and DNN-based estimation vs. $E_b/N_0$ under scenario 3. The proposed method, C-ELM-based estimation, and DNN-based estimation are trained in an offline manner.}
\label{fig.MLoffline}
\end{figure}

First, we assume that the training data can be obtained in an offline manner as assumed in recent literature, e.g., in \cite{8052521,8847452,8933411}. Specifically, a sufficiently large dataset labeled by actual channel responses is provided and the channel conditions in the online estimation stage are the same with the channel models used in the offline training. In this way, the ML-based channel estimation methods can achieve the potential performance of their learning modules. The performance of the proposed method is shown in Fig. \ref{fig.MLoffline} along with two other ML-based estimation methods. These two methods outperform the proposed method, which shows the two other methods have better achievable performance than the proposed method thanks to their non-linear fitting ability.

\begin{figure}
\begin{centering}
\includegraphics[scale=0.55]{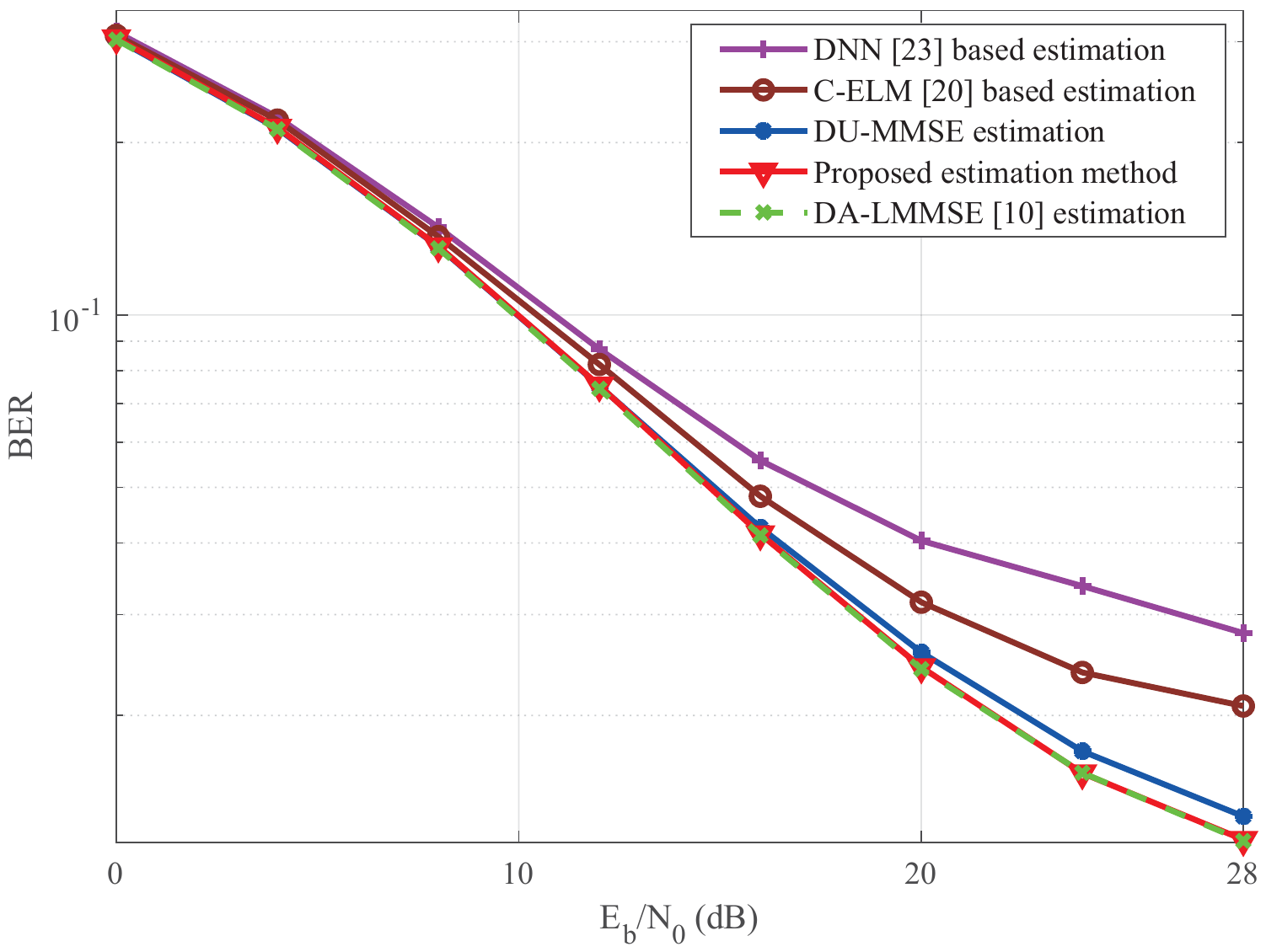}

\end{centering}
\caption{The BER performance of the proposed method, DU-MMSE estimation, DA-LMMSE estimation, C-ELM-based estimation, and DNN-based estimation vs. $E_b/N_0$ under scenario 3. The proposed method, C-ELM-based estimation, and DNN-based estimation employ the online training scheme proposed in this paper.}
\label{fig3}
\end{figure}

Then, we simulate the situation where offline training is not provided and PATDG is applied to generate the training data. The simulation results are given in Fig. \ref{fig3}. It can be seen that although the other two ML-based estimation methods have better achievable performance, the proposed method still outperforms them, which validates the use of the linear learning module. The advantage over the two other ML-based estimation methods stems from its requiring fewer training samples than those methods. In addition, it can be seen that the performance of the proposed method is better than the distortion unaware MMSE (DU-MMSE) channel estimation, where the non-linear distortion is not considered. And its performance approaches the distortion-aware LMMSE (DA-LMMSE) channel estimation \cite{8933050}, where the effective noise variance incorporating non-linear distortion is used. It demonstrates that the proposed method can be trained to approach the optimal linear estimator under a non-linear channel and compensate non-linear distortion to a certain degree.

\subsection{Simulation Results under Changing Scenarios}

Finally, we evaluate the performance of the proposed method under varying channel conditions. We consider the situation where the channel changes within the three scenarios at equal probability when the transmission of the current OFDM frame is ended. In Scenario 2, $\varepsilon_{\rm max} = 0.01$ and $\theta _{\rm min}=-20$. For ML-based channel estimation methods with offline training, it is important for the learning module to work effectively when the channel mismatches the one for training \cite{8052521}. This is because the estimator cannot be trained during transmission although the channel may change. The estimator is usually trained under a proper channel to have robust performance. As the SNR during transmission is unknown, the network is usually trained at a certain SNR value. In the offline training scheme of \cite{8640815} the network is trained at the SNR value of 22 dB.

\begin{figure}
\begin{centering}
\includegraphics[scale=0.49]{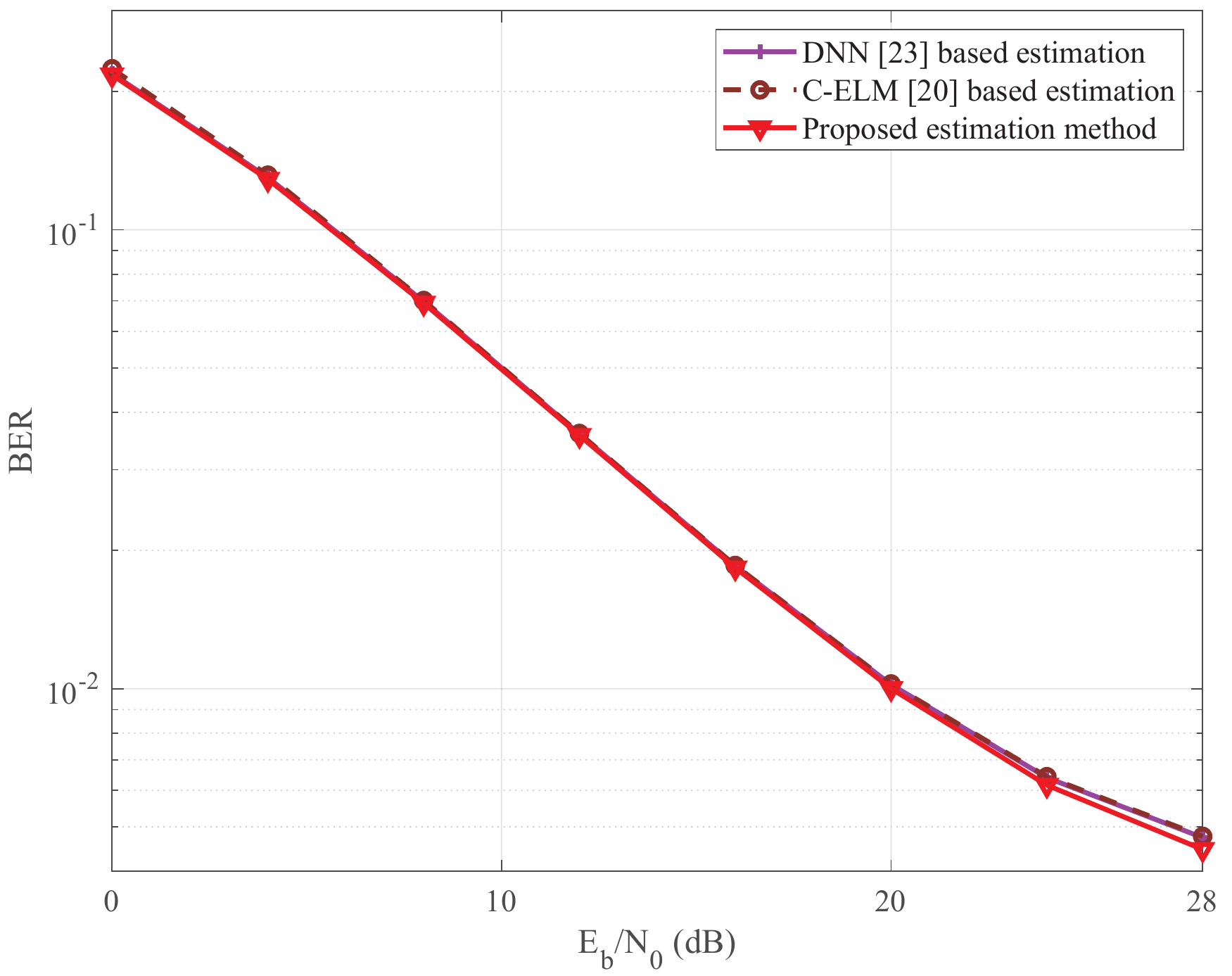}

\end{centering}
\caption{The BER performance of the proposed method, C-ELM-based estimation, and DNN-based estimation vs. $E_b/N_0$ under changing scenarios. The proposed method employs the online training scheme proposed in this paper, while C-ELM-based estimation and DNN-based estimation employ the offline training scheme proposed in \cite{8640815}.}
\label{fig.OFFvsON}
\end{figure}

We compare the proposed method with the two other ML-based channel estimation methods used in the experiment above. Those two methods adopt the offline training scheme of \cite{8640815} and the estimators are trained at the $E_b/N_0$ value of 22 dB under Scenario 3. Scenario 3 is chosen because it shows the best performance among the three scenarios according to our simulation experiments. We simulate their average performance using the Mont Calo method and the results are displayed in Fig. \ref{fig.OFFvsON}. It can be observed that the average performance of the proposed method is better than the two other ML-based channel estimation methods although the two methods have better performance when the channel matches the one for training, as shown in Fig. \ref{fig.MLoffline}. This is because the training data in the proposed method is collected online and it is always trained under the correct channel conditions. It demonstrates the advantage of the proposed online training scheme over the existing ML-based channel estimation with offline training in the situation that the channel changes.

\section{Conclusion}
\label{sec.Conclusion}

In this paper, we propose a low complexity learning-based channel estimation method. Different from the deep learning-based channel estimation, offline training is not required in the proposed method. It is trained online thanks to its fast learning ability and the proposed training data generation schemes, which collect training data online. The proposed training data structure is validated by theoretical analysis and simulations. The simulation results also demonstrate the advantages of the proposed method. First, the proposed channel estimation method is robust to pilot intervals and channel environments. Then, as a data-driven method, the proposed scheme is adaptive to practical impairments like STO and non-linear distortion. Furthermore, since the available training data is limited in the proposed scheme, the employed linear machine learning method has better performance compared to C-ELM and DNN, although these methods have better fitting abilities.

It is worth noting that the proposed channel estimation method fails to compensate CFO, which cannot be addressed by linear operations of the estimator, and with sufficient training data deep learning techniques show higher potential than the employed linear learning module. However, in the proposed scheme, the provided training dataset is small, which hinders the application of deep learning techniques. It would be interesting future work to improve the training data generation method and explore the potential of applying deep learning techniques to the online training scheme.


\ifCLASSOPTIONcaptionsoff
  \newpage
\fi



%
\bibliographystyle{IEEEtran}
\bibliography{TCOM-TPS-20-1301}

%

\begin{IEEEbiography}[{\includegraphics[width=1in,height=1.25in,clip,keepaspectratio]{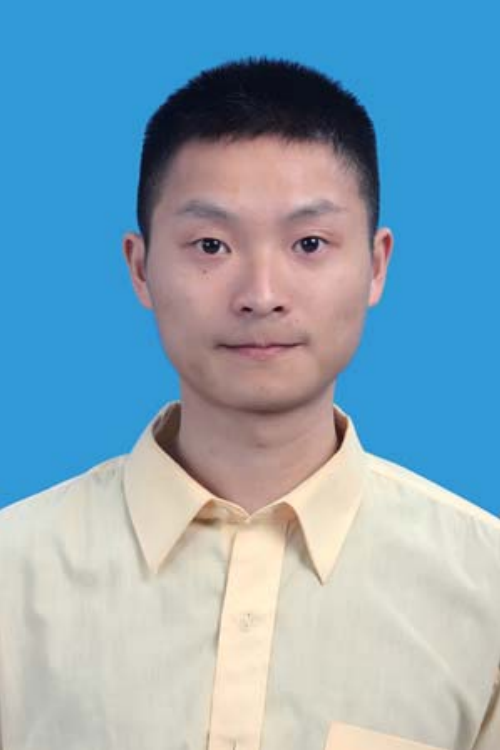}}]{Kai Mei}
received his Master's degree from National University of Defense Technology in 2017. Currently, he is studying at National University of Defense Technology for a Ph.D. degree. He has been a visiting Ph.D. student with the University of Oulu in Finland from 2019 to 2020. His research interests include synchronization and channel estimation in OFDM systems and MIMO-OFDM systems, and machine learning applications in wireless communications.
\end{IEEEbiography}
\begin{IEEEbiography}[{\includegraphics[width=1in,height=1.25in,clip,keepaspectratio]{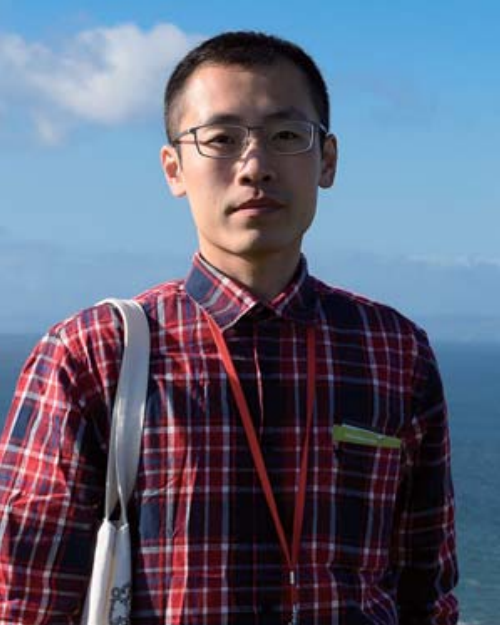}}]{Jun Liu}
received the B.S. degree in optical information science and technology from the South China University of Technology (SCUT), in 2015, and the M.E. degree in communications and information engineering from the National University of Defense Technology (NUDT), Changsha, China, in 2017, where he is currently pursuing the Ph.D. degree with the Department of Cognitive Communications. He is currently a visiting Ph.D. student with the University of Leeds. His current research interests include machine learning with a focus on shallow neural networks applications, signal processing for broadband wireless communication systems, multiple antenna techniques, and wireless channel modeling.
\end{IEEEbiography}
\begin{IEEEbiography}[{\includegraphics[width=1in,height=1.25in,clip,keepaspectratio]
{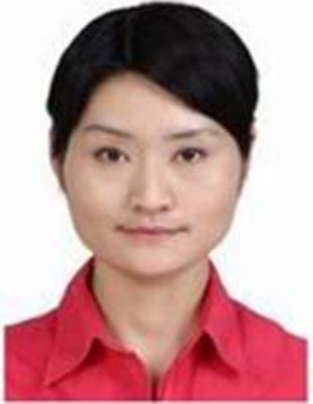}}]{Xiaoying Zhang}
received the M.S. and Ph.D. degrees in Communication Engineering from National University of Defense Technology (NUDT), Changsha, China in 2002 and 2008, respectively. During 2007-2008, she was a visiting scholar in Kyushu University, Japan. Since 2014, she is an associate Professor in NUDT. From Mar. 2017 to Sep. 2017, she was a visiting scholar at the 5G Innovation Centre (5GIC), Institute of Communications (ICS), University of Surrey, U.K. Her main research interests are in wireless communications, including new air interface design, receiver design and iterative signal processing for wireless communication systems.

\end{IEEEbiography}
\begin{IEEEbiography}[{\includegraphics[width=1in,height=1.25in,clip,keepaspectratio]
{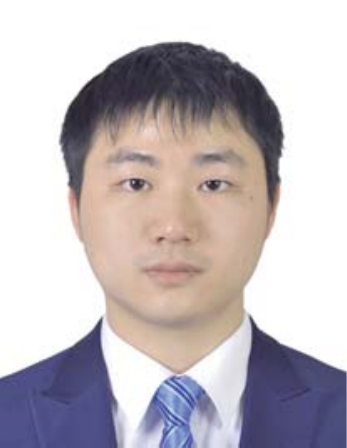}}]{Kuo Cao}
received the B.S. degree in communications engineering and the Ph.D. degree in communications and information systems from the Institution of Communications Engineering, Army Engineering University of PLA, Nanjing, China, in 2013 and 2018, respectively. His research interests are MIMO systems, cooperative communications, and physical layer security of wireless communications.
\end{IEEEbiography}
\begin{IEEEbiography}[{\includegraphics[width=1in,height=1.25in,clip,keepaspectratio]{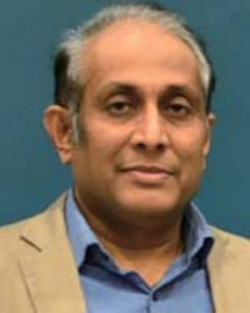}}]{Nandana Rajatheva}
received the B.Sc. (Hons.) degree in electronics and telecommunication engineering from the University of Moratuwa, Sri Lanka, in 1987, ranking first in the graduating class, and the M.Sc. and Ph.D. degrees from the University of Manitoba, Winnipeg, MB, Canada, in 1991 and 1995, respectively. He is currently a Professor with the Centre for Wireless Communications, University of Oulu, Finland. He was a Canadian Commonwealth Scholar during the graduate studies in Manitoba. He held a Professor/Associate Professor positions with the University of Moratuwa and the Asian Institute of Technology, Thailand, from 1995 to 2010. He is currently leading the AI-driven Air Interface design task in Hexa-X EU Project. He has coauthored more than 200 refereed papers published in journals and in conference proceedings. His research interests include physical layer in beyond 5G, machine learning for PHY and MAC, sensing for factory automation and channel coding.
\end{IEEEbiography}
\begin{IEEEbiography}[{\includegraphics[width=1in,height=1.25in,clip,keepaspectratio]{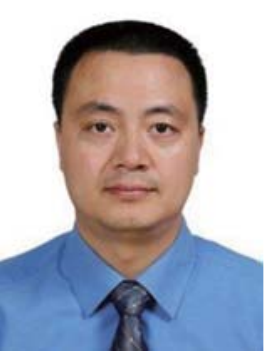}}]{Jibo Wei}
received his BS degree and MS degree from National University of Defense Technology, Changsha, China, in 1989 and 1992, respectively, and the PhD degree from Southeast University, Nanjing, China, in 1998, all in electronic engineering. He is currently a professor of the Department of Communication Engineering of NUDT. His research interests include wireless network protocol and signal processing in communications, cooperative communication, and cognitive network. He is the member of the IEEE Communication Society and also the member of the IEEE VTS. He is a Senior Member of China Institute of Communications and a Senior Member of China Institute of Electronics respectively. He is also an editor of the journal of China Communications.
\end{IEEEbiography}




\end{document}